\documentclass[12pt]{article}
\usepackage{latexsym,amssymb}
\usepackage{amsmath,amsbsy}
\usepackage[all]{xy}
\usepackage{amsopn,amstext}
\usepackage{cite}
\usepackage{amssymb}
\usepackage{rotating}
\usepackage{amsmath}
\usepackage{amsopn,amstext}
\usepackage{pdflscape}
\usepackage{color}
\usepackage[thicklines]{cancel}
\usepackage[colorlinks = true,
linkcolor = blue,
urlcolor  = blue,
citecolor = blue,
anchorcolor = blue]{hyperref}
\usepackage{hyperref}
\usepackage{tikz}
\usetikzlibrary{patterns}
\usepackage{pgfplots}
\usepackage{tikz-3dplot}
\usetikzlibrary{shadows}
\usetikzlibrary{intersections,positioning}
\usetikzlibrary{arrows.meta}
\usepackage{epsfig}
\usepackage{graphicx}
\usepackage{amsthm}
\usepackage{easyReview}
\usepackage{graphicx}
\usepackage{subcaption}
\usepackage{rotating}
\usepackage{bm}

\newtheorem*{remark}{Remark}

\allowdisplaybreaks[4]

\begin{document}



\def\a{\alpha}
\def\b{\beta}
\def\d{\delta}
\def\e{\epsilon}
\def\g{\gamma}
\def\h{\mathfrak{h}}
\def\k{\kappa}
\def\l{\lambda}
\def\o{\omega}
\def\p{\wp}
\def\r{\rho}
\def\t{\tau}
\def\s{\sigma}
\def\z{\zeta}
\def\x{\xi}
\def\V={{{\bf\rm{V}}}}
 \def\A{{\cal{A}}}
 \def\B{{\cal{B}}}
 \def\C{{\cal{C}}}
 \def\D{{\cal{D}}}
\def\G{\Gamma}
\def\K{{\cal{K}}}
\def\O{\Omega}
\def\R{\bar{R}}
\def\T{{\cal{T}}}
\def\L{\Lambda}
\def\f{E_{\tau,\eta}(sl_2)}
\def\E{E_{\tau,\eta}(sl_n)}
\def\Zb{\mathbb{Z}}
\def\Cb{\mathbb{C}}

\def\R{\overline{R}}
\def\ee{\mathrm{e}}

\def\beq{\begin{equation}}
\def\eeq{\end{equation}}
\def\bea{\begin{eqnarray}}
\def\eea{\end{eqnarray}}
\def\ba{\begin{array}}
\def\ea{\end{array}}
\def\no{\nonumber}
\def\le{\langle}
\def\re{\rangle}
\def\lt{\left}
\def\rt{\right}
\global\long\def\mc#1{\mathcal{#1}}
\global\long\def\mf#1{\mathsf{#1}}

\newtheorem{Theorem}{Theorem}
\newtheorem{Definition}{Definition}
\newtheorem{Proposition}{Proposition}
\newtheorem{Lemma}{Lemma}
\newtheorem{Corollary}{Corollary}

\baselineskip=20pt
\newfont{\elevenmib}{cmmib10 scaled\magstep1}
\newcommand{\preprint}{
   \begin{flushleft}
   \end{flushleft}\vspace{-1.3cm}
   \begin{flushright}\normalsize
   \end{flushright}}
\newcommand{\Title}[1]{{\baselineskip=26pt
   \begin{center} \Large \bf #1 \\ \ \\ \end{center}}}

\newcommand{\Author}{\begin{center}
   \large \bf
Guang-Liang Li${}^{a,b}$, Xin Zhang${}^{c}$\footnote{Corresponding author:
xinzhang@iphy.ac.cn}, Junpeng
Cao${}^{b,c,d}$, Wen-Li
Yang${}^{b,e,f,g}\footnote{Corresponding author:
wlyang@nwu.edu.cn}$ and Yupeng Wang${}^{c}$
 \end{center}}

\newcommand{\Address}{\begin{center}
${}^a$ Ministry of Education Key Laboratory for Nonequilibrium
Synthesis and Modulation of Condensed Matter, School of
Physics, Xi'an Jiaotong University, Xi'an 710049, China\\
${}^b$ Peng Huanwu Center for Fundamental Theory, Xi'an 710127, China\\
${}^c$ Beijing National Laboratory for Condensed Matter Physics, Institute of Physics, Chinese Academy of Sciences, Beijing 100190, China\\
${}^d$ School of Physical Sciences, University of Chinese Academy of Sciences, Beijing 100049, China\\
${}^e$ Institute of Modern Physics, Northwest University, Xi'an 710127, China\\
${}^f$ Shaanxi Key Laboratory for Theoretical Physics Frontiers, Xi'an 710127, China\\
${}^g$ Fundamental Discipline Research Center for Quantum Science and Technology of Shaanxi Province, Xi'an 710127, China
\end{center}}

\preprint
\thispagestyle{empty}
\bigskip\bigskip\bigskip

\Title{Integrable Stochastic Processes Associated with the $D_2$ Algebra} \Author

\Address
\vspace{1cm}

\begin{abstract}
 We introduce an integrable stochastic process associated with the $D_2$ quantum group, which can be decomposed into two symmetric simple exclusion processes. We establish the integrability of the model under three types of boundary conditions (periodic, twisted, and open boundaries), and present its exact solution, including the spectrum, eigenstates, and some observables. This integrable model can be generalized to the asymmetric case, decomposing into two asymmetric simple exclusion processes, and its exact solutions are also studied.

\vspace{1truecm} \noindent {\it PACS:} 75.10.Pq, 02.30.Ik,
71.10.Pm

\end{abstract}
\newpage

\tableofcontents 

\section{Introduction}
\setcounter{equation}{0}
Stochastic processes serve as fundamental tools for modeling random phenomena in various scientific disciplines, from mathematics, physics, biology to data science \cite{Van92,Rolski09,Grimmett20}. Among them, integrable stochastic processes represent a special class that possesses exact solvability and rich mathematical structures. The most extensively studied integrable stochastic processes include the symmetric simple exclusion process (SSEP) \cite{Crampe14}, the asymmetric simple exclusion process (ASEP) \cite{Essler96,Derrida98,deGier05,deGier06,Zhang25}, and multi-species ASEP \cite{Ferrari07,Crampe16,Cantini16,Zhang19}.
These integrable stochastic processes form a bridge between probability theory and integrable systems, offering profound insights into non-equilibrium statistical mechanics. Furthermore, these models serve as essential benchmarks for understanding complex real-world stochastic phenomena and inspire the development of analytical and numerical techniques. 

In this work, we construct a novel integrable stochastic process associated with the $D_2$ spin chain \cite{Martins97,DeVega87}, which represents a prototypical quantum integrable system not based on an $A$-type Lie algebra, and has many applications in the high energy, topological and mathematical physics. Our model incorporates four types of particles distinguished by their ``charge" and ``color". Owing to the Lie algebra isomorphism $SO(4) \cong SU(2) \oplus SU(2)$ (often denoted as $D_2 \cong A_1 \oplus A_1$ in Cartan classification), the system can be decomposed into two independent SSEPs (or equivalently, $SU(2)$ XXX spin chains). We demonstrate the integrability of the model under periodic, twisted, and open boundary conditions and derive its exact solution in the form of $T$–$Q$ relation, Bethe state and associated Bethe ansatz equations. Our model can be generalized to the asymmetric case, where the system can be decomposed into two ASEPs. In this scenario, the system may be associated with the $D_2^{(1)}$ spin chain \cite{Lima-Santos:2003,Nepomechie2017,Li2021}. The models introduced in this paper differ from existing multi-component stochastic processes with integrable boundary conditions, thereby enriching the family of exactly solvable stochastic systems.

The paper is structured as follows: In Section \ref{sec:periodic}, Section \ref{sec:twist}, and Section \ref{sec:open}, we study the stochastic process with periodic, twisted, and open boundary conditions, respectively. For each case, we demonstrate the integrability of the model and derive the corresponding exact solutions. We generalize the models to the asymmetric case and present their exact solutions in Section \ref{sec:asep}.
 The paper concludes with a summary and outlook in Section \ref{sec:conclusion}.

\section{Integrable symmetric stochastic process with period boundary conditions}\label{sec:periodic}
\setcounter{equation}{0}
\subsection{Transition matrix}
We introduce a novel stochastic process defined on a one-dimensional lattice, involving four types of particles characterized by their ``charge'' values: $-2$, $-1$, $+1$, and $+2$. These particles can also be classified according to their ``color" property: particles with charges $+1$ and $-1$ share one color, while those with charges $+2$ and $-2$ possess the other color.

\begin{figure}[htbp]
\centering
\begin{subfigure}{0.45\textwidth}

\begin{tikzpicture}
\draw[dashed,color=gray,line width=0.5pt] (0.2,0) --(5.8,0);
\draw[color=black!40,fill] (1,0) circle (.25);
\node at (1,0) {\footnotesize $i$};
\draw[color=blue!40,fill] (3,0) circle (.25);
\node at (3,0) {\footnotesize $j$};
\draw[color=black!40,fill] (5,0) circle (.25);
\node at (5,0) {\footnotesize $k$};
\draw[<->] (1.0,-0.3) arc (-160 : -20 : 1.05);
\coordinate[label=above:$1$] (1) at (2,-0.7);
\draw[<->] (3.0,0.3) arc (160 : 20 : 1.05);
\coordinate[label=below:$1$] (1) at (4,0.7);

\node at (3,-2) {};
\node at (3,2.5) {};
\end{tikzpicture}
\caption{~}\label{fig:1a}
\end{subfigure}
\begin{subfigure}{0.45\textwidth}
\begin{tikzpicture}
\draw[dashed,color=gray,line width=0.5pt] (0.2,0) --(2.8,0);
\draw[color=black!40,fill] (1,0) circle (.25);
\node at (0.98,0) {\footnotesize $-$2};
\draw[color=black!40,fill] (2,0) circle (.25);
\node at (1.98,0) {\footnotesize $+$2};
\draw[dashed,color=gray,line width=0.5pt] (4.2,0) --(6.8,0);
\draw[color=blue!40,fill] (5,0) circle (.25);
\node at (4.98,0) {\footnotesize $-1$};
\draw[color=blue!40,fill] (6,0) circle (.25);
\node at (5.98,0) {\footnotesize $+1$};
\draw[dashed,color=gray,line width=0.5pt] (0.2,-2) --(2.8,-2);
\draw[color=blue!40,fill] (1,-2) circle (.25);
\node at (0.98,-2) {\footnotesize $+1$};
\draw[color=blue!40,fill] (2,-2) circle (.25);
\node at (1.98,-2) {\footnotesize $-1$};
\draw[dashed,color=gray,line width=0.5pt] (4.2,-2) --(6.8,-2);
\draw[color=black!40,fill] (5,-2) circle (.25);
\node at (4.98,-2) {\footnotesize $+2$};
\draw[color=black!40,fill] (6,-2) circle (.25);
\node at (5.98,-2) {\footnotesize $-2$};
\draw[<->] (1.5,0.3) arc (135: 45 : 2.8);
\coordinate[label=below:$1$] (1) at (3.5,0.95);
\draw[<->] (1.5,-2.3) arc (-135: -45 : 2.8);
\coordinate[label=below:$1$] (2) at (3.5,-2.5);
\draw[<->] (1.5,-0.3) arc (150: 210 : 1.45);
\coordinate[label=left:$1$] (1) at (1.2,-1.0);
\draw[<->] (5.5,-0.3) arc (30: -30 : 1.45);
\coordinate[label=right:$1$] (1) at (5.8,-1.0);
\end{tikzpicture}
\caption{~}\label{fig:1b}
\end{subfigure}
\caption{Transition rates of the particles in the model. Here $i, j, k = \pm 1, \pm 2$ denote the ``charge" values of the particles, and different colors in the figure are used to label the particle ``color".}\label{Fig:bulk}
\end{figure}
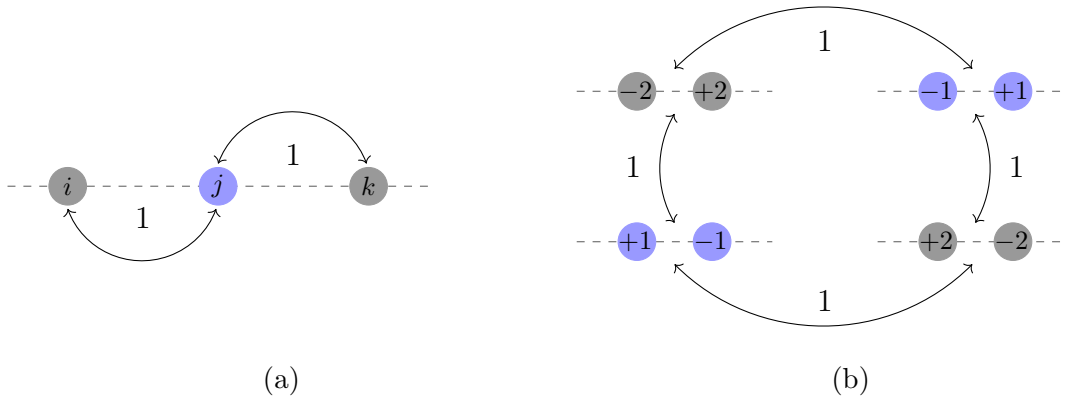

When two particles of different colors occupy adjacent sites, they exchange positions with a symmetric transition rate, analogous to the dynamics in SSEP (see Figure \ref{fig:1a}). 
When particles at adjacent lattice sites have the same color but different charges, they can transform into particles whose color is different from that of the original particles, and the charges of these new particles are distinct from each other (see Figure \ref{fig:1b}).

Let us represent the particles with charges $-2$, $-1$, $+1$, and $+2$ by the following vectors respectively:
$$|-2\rangle=\left(\begin{matrix}
1 \\[-2pt]
0 \\[-2pt]
0 \\[-2pt]
0 \\[-2pt]
\end{matrix}\right), \quad |-1\rangle=\left(\begin{matrix}
0 \\[-2pt]
1 \\[-2pt]
0 \\[-2pt]
0 \\[-2pt]
\end{matrix}\right),\quad |+1\rangle=\left(\begin{matrix}
0 \\[-2pt]
0 \\[-2pt]
1 \\[-2pt]
0 \\[-2pt]
\end{matrix}\right), \quad |+2\rangle=\left(\begin{matrix}
0 \\[-2pt]
0 \\[-2pt]
0 \\[-2pt]
1 \\[-2pt]
\end{matrix}\right). $$ 
Then, the particle transition on sites $k$ and $k+1$ can be described by the following local transition matrix $\mc{M}_{k,k+1}$
\small
\begin{align}
 \mc{M}_{k,k+1}=\left(\begin{array}{cccc|cccc|cccc|cccc}
    0&0&0&0 &0&0&0&0 &0&0&0&0 &0&0&0&0 \\
    0&-1&0&0 &1&0&0&0 &0&0&0&0 &0&0&0&0 \\
    0&0&-1&0 &0&0&0&0 &1&0&0&0 &0&0&0&0 \\
    0&0&0&-2 &0&0&1&0 &0&1&0&0 &0&0&0&0 \\
   \hline 0&1&0&0 &-1&0&0&0 &0&0&0&0 &0&0&0&0 \\
    0&0&0&0 &0&0&0&0 &0&0&0&0 &0&0&0&0 \\
    0&0&0&1 &0&0&-2&0 &0&0&0&0 &1&0&0&0 \\
    0&0&0&0 &0&0&0&-1 &0&0&0&0 &0&1&0&0 \\
   \hline 0&0&1&0 &0&0&0&0 &-1&0&0& 0&0&0&0&0 \\
    0&0&0&1 &0&0&0&0 &0&-2&0&0 &1&0&0&0 \\
    0&0&0&0 &0&0&0&0 &0&0&0&0 &0&0&0&0 \\
    0&0&0&0 &0&0&0&0  &0&0&0&-1 &0&0&1& 0\\
   \hline 0&0&0&0 &0&0&1&0 &0&1&0&0 &-2&0&0&0 \\
    0&0&0&0 &0&0&0&1 &0&0&0&0 &0&-1&0&0 \\
    0&0&0& 0&0&0&0&0 &0&0&0&1 &0&0&-1& 0\\
   0 &0&0&0 &0&0&0&0 &0&0&0&0 &0&0&0&0
    \\
           \end{array}\right),\label{Mkk1}
\end{align} 
\normalsize
which is an operator acting non-trivially in the tensor space  $V_{k}\otimes V_{k+1}$ and as identity on the other tensor spaces.

We consider a one-dimensional lattice  consisting of $N$ sites with periodic boundary conditions. The evolution of this  model is governed by the following master equation
\bea\frac{d}{dt}|\Psi(t)\rangle =\mc{M}|\Psi(t)\rangle. \label{MP}\eea
The transition matrix $\mathcal{M}$ in \eqref{MP} is defined as 
\bea &&\mc{M}=\sum_{k=1}^{N}\mc{M}_{k,k+1},\label{mp}
\eea
where the periodic boundary condition implies  $\mc{M}_{N,N+1}\equiv\mc{M}_{N,1}$.

When the system contains only two types of particles differing in color, it reduces to the SSEP with periodic boundaries. In cases where particles of the same color but different charges are present in the system, the particle numbers are not conserved, nevertheless, the charge conservation law still holds for all possible transitions. 
Let $n_j$ denote the total number of particles with label $j$,
where $j \in \{-2,\,-1,\,+1,\,+2\}$. We see that $\mathcal{Q}_1=n_{+2}+n_{+1}$ and $\mathcal{Q}_2=n_{+2}+n_{-1}$ are two independent conserved charges, each ranging from 
$0$ to $N$.

The matrix $\mathcal{M}$ in \eqref{mp} can be mapped to an integrable $D_2$ spin chain, enabling the derivation of its exact solutions via the Bethe ansatz approach. The integrability and the exact solution of the model is demonstrated in the following sections.

\subsection{Integrability of the model}
\subsubsection{$R$-matrix}
Let's first introduce 
the $R$-matrix related to the $D_2$ algebra \cite{Martins97}
\bea
R_{1,2}(u)=\left(\begin{array}{cccc|cccc|cccc|cccc}
    a&&& &&&& &&&& &&&& \\
    &b&& &g&&& &&&& &&&& \\
    &&b& &&&& &g&&& &&&& \\
    &&&e &&&d& &&d&& &c&&& \\
   \hline &g&& &b&&& &&&& &&&& \\
    &&& &&a&& &&&& &&&& \\
    &&&d &&&e& &&c&& &d&&& \\
    &&& &&&&b &&&& &&g&& \\
   \hline &&g& &&&& &b&&& &&&& \\
    &&&d &&&c& &&e&& &d&&& \\
    &&& &&&& &&&a& &&&& \\
    &&& &&&&  &&&&b &&&g& \\
   \hline &&&c &&&d& &&d&& &e&&& \\
    &&& &&&&g &&&& &&b&& \\
    &&& &&&& &&&&g &&&b& \\
    &&& &&&& &&&& &&&&a \\
           \end{array}\right),\label{rm}
\eea which is defined in the tensor space $V_1\otimes V_2$. In Eq. \eqref{rm}, we use the shorthand notation for the following functions:
\begin{align}
&a(u)=(u+1)^2,\quad  b(u)=u(u+1),\quad  c(u)=1,\no\\ 
&d(u)=u,\quad 
e(u)=u^2,\quad g(u)=u+1.
\end{align}
The $R$-matrix in Eq. (\ref{rm}) possesses the following properties
\begin{align}
{\rm regularity}&:\quad R _{1,2}(0)={P}_{1,2},\nonumber\\
{\rm unitarity}&:\quad R_{1,2}(u)R _{2,1}(-u)=\rho_1(u)\times \mathbb{I},\nonumber\\
{\rm crossing-unitarity}&:\quad R^{t_1} _{1,2}(u)R^{t_1}_{2,1}(-u-2)=\rho_1(u+1)\times \mathbb{I},\nonumber
\end{align}
where $\rho_1(u)=(u^2-1)^2$, $P_{1,2}$ is the
permutation operator with the elements $P^{i,j}_{k,l}=\delta_{i,l}\delta_{j,k}$, $t_i$ denotes the
transposition in $i$-th space, and $R _{2,1}(u)=P_{1,2}R
_{1,2}(u)P_{1,2}=R_{1,2}^{t_1,t_2}(u)$. The $R$-matrix
satisfies the Yang-Baxter equation (YBE)
\begin{eqnarray}
R _{1,2}(u-v)R _{1,3}(u)R_{2,3}(v)=R_{2,3}(v)R _{1,3}(u)R _{1,2}(u-v).\label{YBE}
\end{eqnarray}
\begin{remark}
The $R$-matrix (\ref{rm}) can be obtained by applying a gauge transformation to the $R$ matrix associated with the $D_2$ algebra in Ref. \cite{Martins97}. Let's denote the $R$ matrix in Ref. \cite{Martins97} as $\tilde{R}$, then we have $R=(F\otimes F)\,\tilde{R}\,(F^{-1}\otimes F^{-1})$ with $F={\rm diag}(1,1,-1,1)$.
\end{remark}

\subsubsection{The transfer matrix}
Define the following one-row monodromy matrix
\begin{align}
T_0 (u)&=R _{0,N}(u)R _{0,N-1}(u)\cdots R
_{0,1}(u).\label{T1}
\end{align}
By using the YBE \eqref{YBE} repeatedly, one can prove the RTT relation 
\begin{align}
R_{1,2}(u-v) T_1(u) T_2(v) = T_2(v) T_1(u)R_{1,2}(u-v).\label{RTT}\   
\end{align}
The transfer matrix is constructed as
\bea
t(u)={\rm tr}_0\{T_0(u)\},\label{tup}
\eea
where $0$ is the auxiliary space. 
The RTT relation \eqref{RTT} leads to the fact that the transfer matrices with different spectral parameters commute with each other: $[t(u),\,t(v)]=0$.
The first order derivative of the logarithm of the transfer matrix $t(u)$ in Eq. \eqref{tup} yields the transition matrix $\mc{M}$ in Eq. \eqref{mp}
\begin{align}
\mc{M}&=\left.\frac{\partial \ln t(u)}{\partial
u}\right|_{u=0} -2N\times \mathbb{I}.\label{hhmp}
\end{align}

\subsection{Exact solution of the model}
\subsubsection{Decomposition of the transfer matrix}
The four-dimensional space $V_i$ can be expressed as a tensor product of two  two-dimensional subspaces, i.e., $V_i=V_{\bar i}\otimes V_{\tilde i}$, where $V_{\bar 1}$ and $V_{\tilde 1}$ share the same algebraic structure. We observe that 
the $R$-matrix \eqref{rm} can be
decomposed into \cite{Martins97,Reshetikhin83}
\bea  R_{1,2}(u)=
\mc{R}^{\sigma}_{\bar 1,\bar 2}(u)\otimes \mc{R}^{\tau}_{\tilde 1,\tilde 2}(u) ,\label{rtr} \eea
where $\mathcal{R}^\sigma_{\overline{1},\overline{2}}(u)$ is identical to $\mathcal{R}^\tau_{\widetilde{1},\widetilde{2}}(u)$, both being the six-vertex $R$-matrix
\bea
\mc{R}^{\sigma}_{\bar 1,\bar 2}(u)=\mc{R}^{\tau}_{\tilde1, \tilde 2}(u)=\left(\begin{array}{cccc}
    u+1&0&0&0 \\
   0 &u&1&0 \\
    0&1&u&0 \\
   0 &0&0&u+1
   \end{array}\right).\label{Rxxx}
\eea
\begin{remark}
The decomposition of the $R$-matrix in Eq. (\ref{rtr}) requires further clarification.
One can expand the matrix $\mc{R}^{\sigma}_{\bar 1,\bar 2}(u)$ in the space $V_{\bar 1}\otimes V_{\bar 2}$ as follows
\begin{align}
&\mc{R}^{\sigma}(u)=\sum_{k=1}^4\gamma_k(u)z(k)\otimes z(k),\\ &\gamma_1(u)=\gamma_2(u)=\gamma_3(u)=\tfrac12,\quad \gamma_4(u)=u+\tfrac{1}{2},\no\\
&
z(1)=\sigma_x,\quad z(2)=\sigma_y,\quad z(3)=\sigma_z,\quad z(4)=\mathbb{I}.\no
\end{align}
Consequently, from Eq. (\ref{rtr}), the $R$-matrix $R_{1,2}(u)$ in \eqref{rm} (acting on $V_{1}\otimes V_{2}$) can be written in the form
\begin{align}
R(u)=\sum_{k,k'=1}^4\gamma_k(u)\gamma_{k'}(u)\underbrace{\left(z(k)\otimes z(k')\right)}_{V_1}\otimes \underbrace{\left(z(k)\otimes z(k')\right)}_{V_2}.
\end{align}
\end{remark}
By using the factorization of the $R$-matrix \eqref{rtr}, we find that the transfer matrix defined in \eqref{tup} can be decomposed into the product of two transfer matrices \cite{Martins97,Reshetikhin83,DeVega87}, each corresponding to an XXX spin chain
\bea t(u)={t}_{\sigma}(u)\otimes t_\tau(u). \label{apop} \eea
The sub-transfer matrices $t_{\sigma}(u)$ and $t_{\tau}(u)$ are defined as
\begin{equation}
t_{s}(u)= {\rm tr}_{\bar 0} \{ \mc{T}^{s}_{\bar 0} (u)\}=\left(
\begin{array}{cc}
\mc{A}_s(u) & \mc{B}_s(u) \\
\mc{C}_s(u) & \mc{D}_s(u) 
\end{array}
\right),\quad s=\sigma,\tau,\label{ts-1p}
\end{equation}
where the monodromy matrix $\mc{T}^{s}_{\bar 0}(u)$ reads
\begin{align}
\mc{T}^{s}_{\bar 0}(u)&=\mc{R}^s_{\bar 0,\bar N}(u)\cdots \mc{R}^s_{\bar 0,\bar 2}(u)\mc{R}^s_{\bar 0,\bar 1}(u). \label{Tt111p}
\end{align}
It should be noted that the physical spaces associated with $t_{\sigma}(u)$ and  $t_{\tau}(u)$ in Eq. \eqref{ts-1p} are different, and their tensor product constitutes the physical space of the $ D_2$ spin chain.

The monodromy matrix $\mc{T}^{s}_{\bar 0}(u)$ satisfies the RTT relation
\bea
&& \mc{R}_{\bar 1,\bar 2}^s(u-v) \mc{T}^{s}_{\bar 1}(u)  \mc{T}^{s}_{\bar 2}(v)= \mc{T}^{s}_{\bar 2}(v) \mc{T}^{s}_{\bar 1}(u) \mc{R}^s_{\bar 1,\bar 2}(u-v),\label{ai2-1p}
\eea
which leads to
\bea
[t_{s}(u),\, t_{s'}(v)]=0,\quad s,s'=\sigma,\tau.
\eea
The first order derivative of $\ln t_{\sigma,\tau}(u)$ gives the Markov transition matrix of SSEP with periodic boundary (or equivalently, the Hamiltonian of a periodic XXX spin chain)
\begin{align}
\mc{M}_{\sigma}&=\left.\frac{\partial \ln t_\sigma(u)}{\partial
u}\right|_{u=0} -N\times \mathbb{I}\no\\
&=\frac12\sum_{j=1}^{N}(\sigma^x_j\sigma^x_{j+1}+\sigma^y_j\sigma^y_{j+1}+
\sigma^z_j\sigma^z_{j+1}-\mathbb{I}),\\
\mc{M}_{\tau}&=\mc{M}_{\sigma}|_{
\{\sigma_j^\alpha\}\rightarrow \{\tau_j^\alpha,  \}}, \quad \alpha=x,y,z.\end{align} 
Consequently, the transition matrix in \eqref{mp} is a direct sum of the transition matrices of two SSEPs with periodic boundaries 
\bea
\mc{M}=\mc{M}_{\sigma}\oplus
\mc{M}_{\tau}. \label{apop1p}
\eea

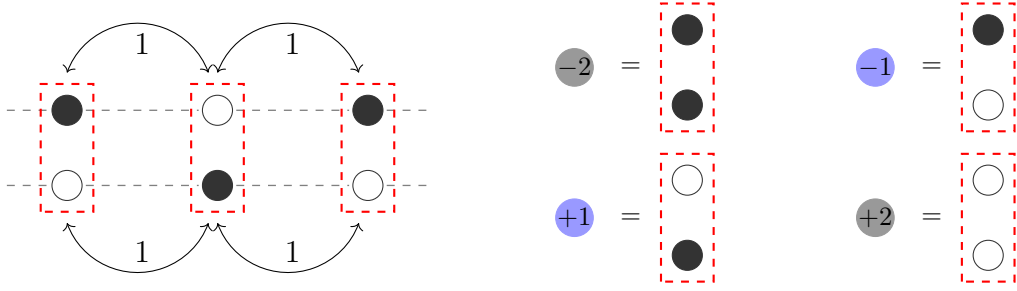
\begin{figure}[htbp]
\centering
\begin{tikzpicture}
\draw[dashed,color=gray,line width=0.5pt] (0.2,0) --(0.75,0);
\draw[dashed,color=gray,line width=0.5pt] (1.25,0) --(2.75,0);
\draw[dashed,color=gray,line width=0.5pt] (3.25,0) --(4.75,0);
\draw[dashed,color=gray,line width=0.5pt] (5.25,0) --(5.8,0);
\draw[color=black!80,fill] (1,0) circle (.2);
\draw[color=black!80] (3,0) circle (.2);
\draw[color=black!80,fill] (5,0) circle (.2);
\draw[<->] (1.0,0.5) arc (160 : 20 : 1.0);
\coordinate[label=above:$1$] (1) at (2,0.6);
\draw[<->] (3.0,0.5) arc (160 : 20 : 1.0);
\coordinate[label=above:$1$] (1) at (4,0.6);
\draw[dashed,color=gray,line width=0.5pt] (0.2,-1) --(0.75,-1);
\draw[dashed,color=gray,line width=0.5pt] (1.25,-1) --(2.75,-1);
\draw[dashed,color=gray,line width=0.5pt] (3.25,-1) --(4.75,-1);
\draw[dashed,color=gray,line width=0.5pt] (5.25,-1) --(5.8,-1);
\draw[color=black!80] (1,-1) circle (.2);
\draw[color=black!80,fill] (3,-1) circle (.2);
\draw[color=black!80] (5,-1) circle (.2);

\draw[<->] (1.0,-1.5) arc (-160 : -20 : 1.0);
\coordinate[label=below:$1$] (1) at (2,-1.6);
\draw[<->] (3.0,-1.5) arc (-160 : -20 : 1.0);
\coordinate[label=below:$1$] (1) at (4,-1.6);
\draw[red,dashed,thick] (0.65,-1.35) rectangle (1.35,0.35);
\draw[red,dashed,thick] (2.65,-1.35) rectangle (3.35,0.35);
\draw[red,dashed,thick] (4.65,-1.35) rectangle (5.35,0.35);
\end{tikzpicture}
\hspace{1.3cm}
\begin{tikzpicture}
\draw[color=black!40,fill] (-0.5,-0.5) circle (.25);
\node at (-0.5,-0.5) {\footnotesize $-2$};
\node at (0.25,-0.5) {\footnotesize $=$};
\draw[red,dashed,thick] (0.65,-1.35) rectangle (1.35,0.35);
\draw[color=black!80,fill] (1,0) circle (.2);
\draw[color=black!80,fill] (1,-1) circle (.2);
\draw[color=blue!40,fill] (3.5,-0.5) circle (.25);
\node at (3.5,-0.5) {\footnotesize $-1$};
\node at (4.25,-0.5) {\footnotesize $=$};
\draw[red,dashed,thick] (4.65,-1.35) rectangle (5.35,0.35);
\draw[color=black!80,fill] (5,0) circle (.2);
\draw[color=black!80] (5,-1) circle (.2);
\draw[color=blue!40,fill] (-0.5,-2.5) circle (.25);
\node at (-0.5,-2.5) {\footnotesize $+1$};
\node at (0.25,-2.5) {\footnotesize $=$};
\draw[red,dashed,thick] (0.65,-3.35) rectangle (1.35,0.35-2);
\draw[color=black!80] (1,0-2) circle (.2);
\draw[color=black!80,fill] (1,-1-2) circle (.2);
\draw[color=black!40,fill] (3.5,-2.5) circle (.25);
\node at (3.5,-2.5) {\footnotesize $+2$};
\node at (4.25,-2.5) {\footnotesize $=$};
\draw[red,dashed,thick] (4.65,-1.35-2) rectangle (5.35,0.35-2);
\draw[color=black!80] (5,-2) circle (.2);
\draw[color=black!80] (5,-3) circle (.2);
\end{tikzpicture}
\caption{Decomposition of the stochastic process. Here, the solid and empty circles denote a particle and a hole, respectively.}\label{Fig2}
\end{figure}

\begin{figure}
\centering
\begin{tikzpicture}
\draw[dashed,color=gray,line width=0.5pt] (0.2,0) --(2.8,0);
\draw[color=black!40,fill] (1,0) circle (.25);
\node at (0.98,0) {\footnotesize $-$2};
\draw[color=black!40,fill] (2,0) circle (.25);
\node at (1.98,0) {\footnotesize $+$2};
\draw[dashed,color=gray,line width=0.5pt] (4.2,0) --(6.8,0);
\draw[color=blue!40,fill] (5,0) circle (.25);
\node at (4.98,0) {\footnotesize $-1$};
\draw[color=blue!40,fill] (6,0) circle (.25);
\node at (5.98,0) {\footnotesize $+1$};
\draw[->] (1.5,0.3) arc (135: 45 : 2.8);
\coordinate[label=below:$1$] (1) at (3.5,0.95);

\node at (3,-1) {\footnotesize ~~~};
\end{tikzpicture}
\hspace{0.5cm}
\begin{tikzpicture}
\draw[dashed,color=gray,line width=0.5pt] (0.2,0) --(0.75,0);
\draw[dashed,color=gray,line width=0.5pt] (1.25,0) --(2.75,0);
\draw[dashed,color=gray,line width=0.5pt] (3.25,0) --(3.75,0);
\draw[color=black!80,fill] (1,0) circle (.2);
\draw[color=black!80] (3,0) circle (.2);

\draw[dashed,color=gray,line width=0.5pt] (0.2,-1) --(0.75,-1);
\draw[dashed,color=gray,line width=0.5pt] (1.25,-1) --(2.75,-1);
\draw[dashed,color=gray,line width=0.5pt] (3.25,-1) --(3.75,-1);
\draw[color=black!80,fill] (1,-1) circle (.2);
\draw[color=black!80] (3,-1) circle (.2);
\draw[<->] (1.0,-1.5) arc (-160 : -20 : 1.0);
\coordinate[label=below:$1$] (1) at (2,-1.6);
\draw[red,dashed,thick] (0.65,-1.35) rectangle (1.35,0.35);
\draw[red,dashed,thick] (2.65,-1.35) rectangle (3.35,0.35);
\node at (-1,-1) {\footnotesize $=$};;
\end{tikzpicture}
\caption{Illustration of the transition process. As an example, we consider the transition process from $|-2\rangle \otimes |+2\rangle$ to $|-1\rangle \otimes |+1\rangle$.}\label{Fig3}
\end{figure}
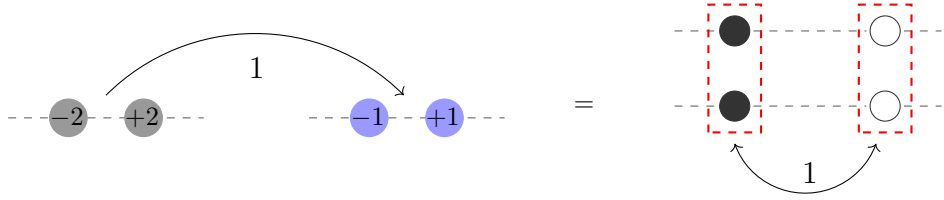

In another physical framework, the stochastic process defined in Eq. (\ref{Mkk1}) can be decoupled into two \textit{independent} lines. On each line, particles hop symmetrically to adjacent sites, which constitutes SSEP. An illustration of the decomposition is provided in Figures \ref{Fig2} and \ref{Fig3}.

\subsubsection{Exact solutions}
\paragraph{$T$-$Q$ relation}
Based on the decomposition in Eq. (\ref{apop}), the eigenvalue of the transfer matrix 
$t(u)$ in \eqref{tup}, denoted by $\Lambda(u)$, can be derived as follows \bea  \Lambda(u)=
\Lambda_{\sigma}(u)\Lambda_{\tau}(u),
\eea
where $\Lambda_{\sigma}(u)$ and $\Lambda_{\tau}(u)$ are the eigenvalues of $t_\sigma(u)$ and $t_\tau(u)$, respectively. The function $\Lambda_\sigma(u)$ can be parameterized by the following  homogeneous $T$-$Q$ relation \cite{Korepin97}
\begin{align} \Lambda_{\sigma}(u)=(u+1)^{N}\,\frac{ Q(u-1)}{ Q(u)}+u^N\,\frac{ Q(u+1)}{Q(u)}, \label{tserp}
\end{align}
where
\bea
Q(u)=\prod_{l=1}^{M}(u-\mu_l),\quad 0\leq M\leq N.\eea
The Bethe roots $\{\mu_l\}$ in Eq. \eqref{tserp} should satisfy the Bethe ansatz equations (BAEs)
\bea
\left[\frac{\mu_{k}}{\mu_{k}+1}\right]^{N}=
 \prod_{l\ne k}^{M}\frac{(\mu_k-\mu_l-1)}
 {(\mu_k-\mu_l+1)},\,
 \qquad k=1,\ldots, M. \label{BA012p}
 \eea

Function $\Lambda_{\tau}(u)$ can be obtained directly from $\Lambda_{\sigma}(u)$ with the following substitution 
\bea
\{ \mu_j,M\}\rightarrow \{\nu_j,M'\}.\eea
Then the eigenvalue of
transition matrix $\cal{M}$ in Eq. \eqref{mp} in terms of the Bethe roots reads
\begin{align} 
E=&\,\left.\frac{\partial \ln (\Lambda_{\sigma}(u)\Lambda_{\tau}(u))}{\partial u}\right|_{u=0}-2N\no\\
=&\sum_{k=1}^M\frac{1}{\mu_k(\mu_k+1)}+\sum_{k=1}^{M'}\frac{1}{\nu_k(\nu_k+1)}.\label{energyp}
\end{align}

\paragraph{Bethe state}

The eigenstate of the transfer matrix $t(u)$ is also factorized, with the following form 
\begin{align}
&|\mu_1,\ldots,\mu_M\rangle_\sigma\otimes |\nu_1,\ldots,\nu_{M'}\rangle_\tau,\\
&\langle\mu_1,\ldots,\mu_M|_\sigma\otimes \langle\nu_1,\ldots,\nu_{M'}|_\tau,
\end{align}
where
\begin{align}
&|u_1,\ldots,u_m\rangle_s=\prod_{k=1}^m\mc{B}_s(u_k)\binom{1}{0}^{\otimes N},\quad
{\rm or}\quad |u_1,\ldots,u_m\rangle_s=\prod_{k=1}^m\mc{C}_s(u_k)\binom{0}{1}^{\otimes N},\label{Bethe:state:xxx}\\
&\langle u_1,\ldots,u_m|_s=(1,\,0)^{\otimes N}\prod_{k=1}^m\mc{C}_s(u_k),\quad
{\rm or}\quad \langle u_1,\ldots,u_m|_s=(0,\,1)^{\otimes N}\prod_{k=1}^m\mc{B}_s(u_k).
\end{align}

The transfer matrices $t_\tau(u)$ and $t_\sigma(u)$, while having identical functional forms, act on independent spaces and are thus mutually independent, which allows us to get all the $2^N \times 2^N \equiv 4^N$ eigenvalues of $t(u)$ (and the transition matrix $\cal{M}$). Numeric results are shown in Table.\ref{Tab1}. The eigenvalues of $\cal{M}$ given by Eq. \eqref{energyp} are consistent with those obtained from exact diagonalization of the transition matrix.

\begin{figure}[htbp]
\centering
\includegraphics[width=0.45\textwidth]{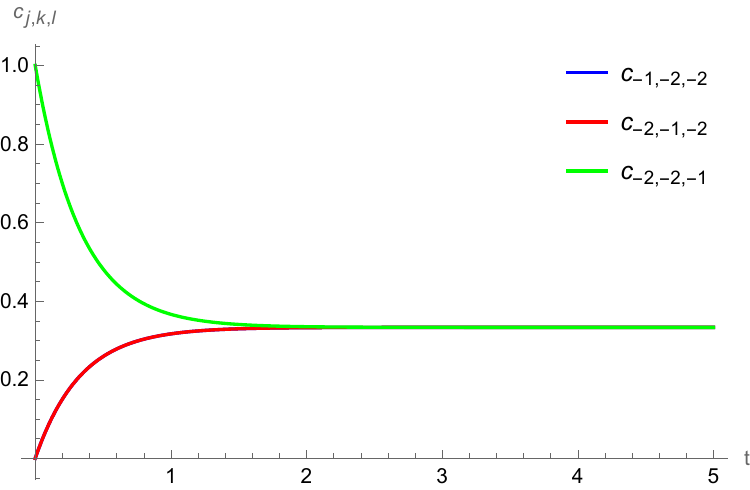}~~~~~\includegraphics[width=0.45\textwidth]{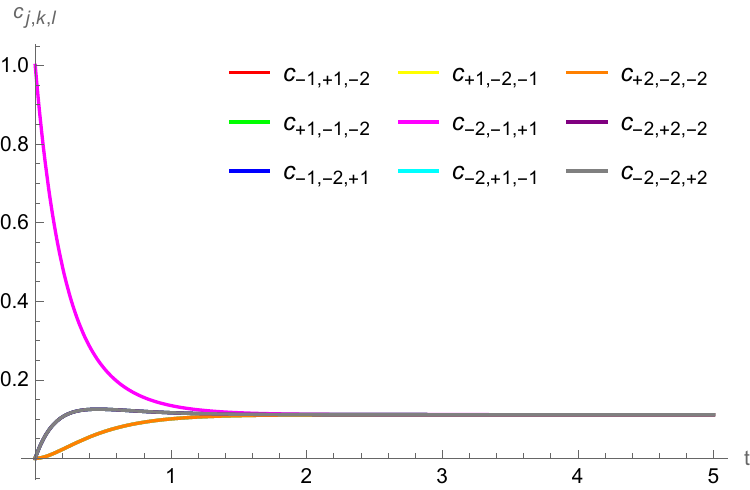}
\caption{{\bf Left panel}: The evolution of $|-2\rangle\otimes|-2\rangle\otimes|-1\rangle$ with $N=3$. We observe that $c_{-1,-2,-2}=c_{-2,-1,-2}$ and $c_{j,k,l}\to \frac13$ in the limit $t\to\infty$. {\bf Right panel:} The evolution of $|-2\rangle\otimes|-1\rangle\otimes|+1\rangle$ with $N=3$. Here, $c_{-1,+1,-2}=c_{+1,-2,-1}=c_{-2,+1,-1}=c_{+2,-2,-2}$, $c_{+1,-1,-2}=c_{-1,-2,+1}=c_{-2,+2,-2}=c_{-2,-2,+2}$ and $c_{j,k,l}\to\frac19$ in the limit $t\to\infty$.}\label{fig:evolution}
\end{figure}

\paragraph{Non-equilibrium steady state}
In stochastic processes, the transition matrix has eigenvalues with non-positive real parts, with $E=0$ corresponding to the non-equilibrium steady state.
Due to the ${SU}(2)$ symmetry of the XXX spin chain, the model possesses $(N+1)^2$  steady states, which correspond to configurations where all the Bethe roots take infinite values
$$\mu_1, \ldots, \mu_M \to \infty, \quad \nu_1, \ldots, \nu_{M'} \to \infty.$$
In the limit $u\to \infty$, the rescaled operator $\mc{B}_s(u)/u^{N-1}$ converges to the total spin‑lowering operator
\begin{align}
\lim_{u\to\infty}\frac{\mc{B}_s(u)}{u^{N-1}}=\sum_{n=1}^N\sigma_n^-,\quad s=\sigma,\tau.
\end{align}
Consequently, the steady states of the model read
\begin{align}
|\Psi_{m,n}\rangle &=|\psi_m\rangle_\sigma\otimes|\psi_n\rangle_\tau,\quad m,n=0,1,\ldots,N,\\
|\psi_m\rangle_\sigma&=\frac{1}{\binom{N}{m}}\sum_{1\leq n_1<n_2\cdots<n_m\leq N}\sigma_{n_1}^-\cdots\sigma_{n_m}^-\binom{1}{0}^{\otimes N},\\
|\psi_m\rangle_\tau&=\frac{1}{\binom{N}{m}}\sum_{1\leq n_1<n_2\cdots<n_m\leq N}\tau_{n_1}^-\cdots\tau_{n_m}^-\binom{1}{0}^{\otimes N}.
\end{align}
Since $t(u)$ (and ${\mathcal{M}}$) is Hermitian, we can easily get the corresponding left steady states.
Conservation of $\mathcal{Q}_{1,2}$ restricts the evolution of an initial state $|\Phi\rangle$ with charges $\mathcal{Q}_{1,2}$ to superpositions of states with identical charges
\begin{align}
\ee^{\mathcal{M}\,t}|\Phi\rangle_{\mathcal{Q}_1,\mathcal{Q}_2}=\sum_{\substack{j_1,\ldots,j_N=\pm2,\pm 1\\ \mbox{fixed}\,\mathcal{Q}_1,\mathcal{Q}_2}}c_{j_1,\ldots,j_N}(t)\bigotimes_{n=1}^N|j_n\rangle,
\end{align} 
where  $c_{j_1,\ldots,j_N}(t)$ is interpreted as the probability of observing the specific particle arrangement $j_1,\ldots,j_N$  at time $t$.  In the long-time limit  \( t \to \infty \)
and the system relaxes to the corresponding steady state
\begin{align}
\lim_{t\to \infty}\ee^{\mathcal{M}\,t}|\Phi\rangle_{\mathcal{Q}_1,\mathcal{Q}_2}=|\Psi_{\mathcal{Q}_1,\mathcal{Q}_2}\rangle,\quad \lim_{t\to \infty}c_{j_1,\ldots,j_N}(t)=\frac{1}{\binom{N}{\mathcal{Q}_1}\binom{N}{\mathcal{Q}_2}}.
\end{align}
As an example, Figure \ref{fig:evolution} shows the relaxation behavior of specific initial states in the $N=3$ case.
  
We now consider the correlation functions in the steady state. The density operator $\hat{n}_{j,k}$, where $j = 1,\dots,N$ and $k = \pm 1, \pm 2$, is also factorized
\begin{align}
\hat{n}_{j,-2}=(\hat{n}_{j,\uparrow})_\sigma\otimes (\hat{n}_{j,\uparrow})_\tau,\quad \hat{n}_{j,-1}=(\hat{n}_{j,\uparrow})_\sigma\otimes (\hat{n}_{j,\downarrow})_\tau,\no\\
\hat{n}_{j,+1}=(\hat{n}_{j,\downarrow})_\sigma\otimes (\hat{n}_{j,\uparrow})_\tau,\quad \hat{n}_{j,+2}=(\hat{n}_{j,\downarrow})_\sigma\otimes (\hat{n}_{j,\downarrow})_\tau,
\end{align}
where $\hat n_{\uparrow}={\rm diag}\{1,\,0\}$ and $\hat n_{\downarrow}={\rm diag}\{0,\,1\}$.
We first review several well-known correlation functions of the periodic SSEP
\begin{align}
\begin{aligned}
&\frac{_s\langle\psi_{m}|(\hat{n}_{j,\uparrow})_s|\psi_{m}\rangle_s}{_s\langle\psi_{m}|\psi_{m}\rangle_s}=\frac{N-m}{N},\quad \frac{_s\langle\psi_{m}|(\hat{n}_{j,\uparrow})_s|\psi_{m}\rangle_s}{_s\langle\psi_{m}|\psi_{m}\rangle_s}=\frac{m}{N},\\
&\frac{_s\langle\psi_{m}|(\hat{n}_{j_1,\uparrow})_s(\hat{n}_{j_2,\uparrow})_s|\psi_{m}\rangle_s}{_s\langle\psi_{m}|\psi_{m}\rangle_s}=\frac{(N-m)(N-m-1)}{N(N-1)},\quad j_1\neq j_2,\\ &\frac{_s\langle\psi_{m}|(\hat{n}_{j_1,\uparrow})_s(\hat{n}_{j_2,\uparrow})_s|\psi_{m}\rangle_s}{_s\langle\psi_{m}|\psi_{m}\rangle_s}=\frac{m(m-1)}{N(N-1)},\quad j_1\neq j_2,\\
&\frac{_s\langle\psi_{m}|(\hat{n}_{j_1,\uparrow})_s(\hat{n}_{j_2,\downarrow})_s|\psi_{m}\rangle_s}{_s\langle\psi_{m}|\psi_{m}\rangle_s}=\frac{m(N-m)}{N(N-1)},\quad j_1\neq j_2,\quad \dots
\end{aligned}
\end{align}
Due to the factorized structure of both the operator and the steady state, the correlation functions of the steady state \(|\Psi_{m,n}\rangle\) can be obtained in the following straightforward way
\begin{align}
\begin{aligned}
\frac{\langle\Psi_{m,n}|\hat{n}_{j,-2}|\Psi_{m,n}\rangle}{\langle\Psi_{m,n}|\Psi_{m,n}\rangle}&=\frac{_\sigma\langle\psi_{m}|(\hat{n}_{j,\uparrow})_\sigma|\psi_{m}\rangle_\sigma}{_\sigma\langle\psi_{m}|\psi_{m}\rangle_\sigma}\,\frac{_\tau\langle\psi_{n}|(\hat{n}_{j,\uparrow})_\tau|\psi_{n}\rangle_\tau}{_\tau\langle\psi_{n}|\psi_{n}\rangle_\tau}\\
\frac{\langle\Psi_{m,n}|\hat{n}_{j_1,-2}\hat{n}_{j_2,-2}|\Psi_{m,n}\rangle}{\langle\Psi_{m,n}|\Psi_{m,n}\rangle}&=\frac{_\sigma\langle\psi_{m}|(\hat{n}_{j_1,\uparrow})_\sigma(\hat{n}_{j_2,\uparrow})_\sigma|\psi_{m}\rangle_\sigma}{_\sigma\langle\psi_{m}|\psi_{m}\rangle_\sigma}\,\frac{_\tau\langle\psi_{n}|(\hat{n}_{j_1,\uparrow})_\tau(\hat{n}_{j_2,\uparrow})_\tau|\psi_{n}\rangle_\tau}{_\tau\langle\psi_{n}|\psi_{n}\rangle_\tau},\\
&\hspace{3,0cm}\dots
\end{aligned}
\end{align}

\begin{figure}[htbp]
\centering
\includegraphics[width=0.45\textwidth]{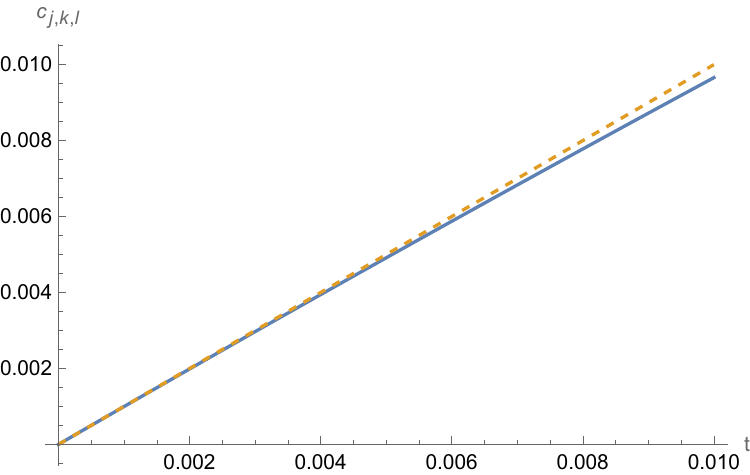}~~~~~\includegraphics[width=0.45\textwidth]{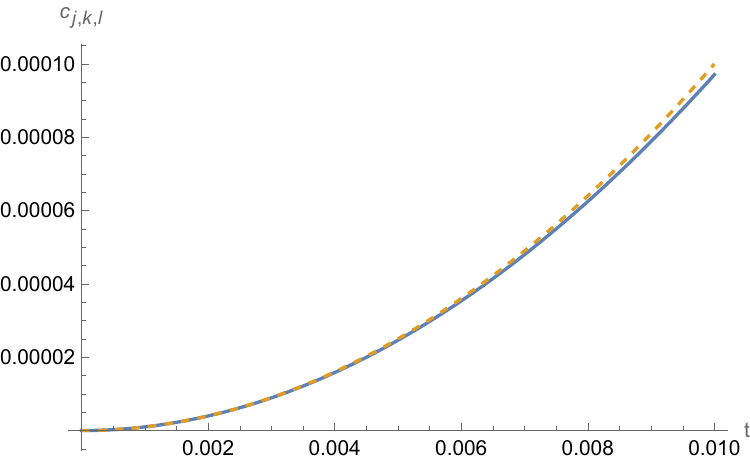}
\caption{Dependence of the expansion coefficient $c_{j_1,\ldots,j_N}$ on time $t$ in the regime $t\ll 1$. The initial state is $|-2\rangle\otimes|-1\rangle\otimes|+1\rangle$ with $N=3$. \textbf{Left panel:} Time evolution of the coefficient $c_{-2,-2,+2}$; the dashed line corresponds to $y = t$. \textbf{Right panel:} Time evolution of the coefficient $c_{-1,+1,-2}$; the dashed line corresponds to $y = t^2$.}\label{fig:coefficient}
\end{figure}

In the regime $t\ll 1$, the relaxation of a quantum state can be also understood.  Let us first expand the operator $\ee^{\mathcal M t}$
\begin{align}
\ee^{\mathcal M t}=\mathbb{I}+\mathcal M t+\frac{\mathcal M^2t^2}{2!}+\frac{\mathcal M^3t^3}{3!}+\dots,\quad t<1.
\end{align}  
Therefore, if a state $\bigotimes_{n=1}^N|j_n\rangle$ is accessible from the initial state via $m$ transitions, its coefficient $c_{j_1,\ldots,j_M}(t)$ in the short-time expansion is of order $t^m$, see Figure \ref{fig:coefficient} for an example.
\begin{table}[htbp]
    \centering
    \begin{tabular}{|c|c|}
    \hline
      $\mu_1/\nu_1$   &  $\mu_2/\nu_2$ \\
      \hline 
       --- & --- \\
      $-$0.5000   & --- \\
      $-$0.5000+0.5000i & --- \\
      $-$0.5000$-$0.5000i & --- \\
      $\infty$ & --- \\
      $-$0.5000   & $\infty$ \\
     $-$0.5000+0.5000i & $\infty$ \\
     $-$0.5000-0.5000i & $\infty$ \\
     $\infty$ & $\infty$ \\
     $-$0.5000+0.2887i & $-$0.5000-0.2887i\\
     $-$1.0000 & 0.0000 \\
    \hline
    \end{tabular}
    \caption{Numeric solutions of BAEs \eqref{BA012p} with $N=4$.}
    \label{Tab1}
\end{table}

\section{Integrable symmetric stochastic process with twisted boundary condition}\label{sec:twist}
\setcounter{equation}{0}

\subsection{Transition matrix}
In this section, we introduce another stochastic process under twisted boundary conditions, whose time evolution is governed by the following master equation:
\bea\frac{d}{dt}|\Psi(t)\rangle =\overline{\mc{M}}|\Psi(t)\rangle. \label{MT}\eea
The transition matrix $\mc{\overline M}$ in Eq. \eqref{MT} reads
\bea &&\mc{\overline  M}=\sum_{k=1}^{N-1}\mc{M}_{k,k+1}+\mc{\overline M}_{N,1},\label{mt}
\eea
where $\mc{M}_{k,k+1}$ is defined in \eqref{Mkk1}, and the additional term $\mc{\overline M}_{N,1}$ in Eq. \eqref{mt} encodes the twisted boundary condition, taking the form
\small
\begin{align}
 \mc{\overline M}_{N,1}=\left(\begin{array}{cccc|cccc|cccc|cccc}
    -2&0&0&0 &0&1&0&0 &0&0&1&0 &0&0&0&0 \\
    0&-1&0&0 &0&0&0&0 &0&0&0&1 &0&0&0&0 \\
    0&0&-1&0 &0&0&0&1 &0&0&0&0 &0&0&0&0 \\
    0&0&0&0 &0&0&0&0 &0&0&0&0 &0&0&0&0 \\
   \hline 0&0&0&0 &-1&0&0&0 &0&0&0&0 &0&0&1&0 \\
    1&0&0&0 &0&-2&0&0 &0&0&0&0 &0&0&0&1 \\
    0&0&0&0 &0&0&0&0 &0&0&0&0 &0&0&0&0 \\
    0&0&1&0 &0&0&0&-1 &0&0&0&0 &0&0&0&0 \\
   \hline 0&0&0&0 &0&0&0&0 &-1&0&0& 0&0&1&0&0 \\
    0&0&0&0 &0&0&0&0 &0&0&0&0 &0&0&0&0 \\
    1&0&0&0 &0&0&0&0 &0&0&-2&0 &0&0&0&1 \\
    0&1&0&0 &0&0&0&0  &0&0&0&-1 &0&0&0& 0\\
   \hline 0&0&0&0 &0&0&0&0 &0&0&0&0 &0&0&0&0 \\
    0&0&0&0 &0&0&0&0 &1&0&0&0 &0&-1&0&0 \\
    0&0&0& 0&1&0&0&0 &0&0&0&0 &0&0&-1& 0\\
   0 &0&0&0 &0&1&0&0 &0&0&1&0 &0&0&0&-2
    \\
           \end{array}\right).\label{Mn1t}
\end{align}
\normalsize

\begin{figure}[htbp]
\centering
\scalebox{0.75}{
\begin{tikzpicture}
\draw[dashed,color=gray,line width=0.5pt] (0.2,0) --(2.8,0);
\draw[color=black!40,fill] (1,0) circle (.25);
\node at (0.98,0) {\footnotesize $-2$};
\draw[color=blue!40,fill] (2,0) circle (.25);
\node at (1.98,0) {\footnotesize $-1$};
\draw[dashed,color=gray,line width=0.5pt] (4.2,0) --(6.8,0);
\draw[color=blue!40,fill] (5,0) circle (.25);
\node at (4.98,0) {\footnotesize $+1$};
\draw[color=black!40,fill] (6,0) circle (.25);
\node at (5.98,0) {\footnotesize $+2$};
\draw[dashed,color=gray,line width=0.5pt] (0.2,-2) --(2.8,-2);
\draw[color=black!40,fill] (1,-2) circle (.25);
\node at (0.98,-2) {\footnotesize $+2$};
\draw[color=blue!40,fill] (2,-2) circle (.25);
\node at (1.98,-2) {\footnotesize $+1$};
\draw[dashed,color=gray,line width=0.5pt] (4.2,-2) --(6.8,-2);
\draw[color=blue!40,fill] (5,-2) circle (.25);
\node at (4.98,-2) {\footnotesize $-1$};
\draw[color=black!40,fill] (6,-2) circle (.25);
\node at (5.98,-2) {\footnotesize $-2$};
\draw[<->] (1.5,0.3) arc (135: 45 : 2.8);
\coordinate[label=below:$1$] (1) at (3.5,0.95);
\draw[<->] (1.5,-2.3) arc (-135: -45 : 2.8);
\coordinate[label=below:$1$] (2) at (3.5,-2.5);
\end{tikzpicture}
\quad 
\begin{tikzpicture}
\draw[dashed,color=gray,line width=0.5pt] (0.2,0) --(2.8,0);
\draw[color=black!40,fill] (1,0) circle (.25);
\node at (0.98,0) {\footnotesize $-2$};
\draw[color=blue!40,fill] (2,0) circle (.25);
\node at (1.98,0) {\footnotesize $+1$};
\draw[dashed,color=gray,line width=0.5pt] (4.2,0) --(6.8,0);
\draw[color=blue!40,fill] (5,0) circle (.25);
\node at (4.98,0) {\footnotesize $-1$};
\draw[color=black!40,fill] (6,0) circle (.25);
\node at (5.98,0) {\footnotesize $+2$};
\draw[dashed,color=gray,line width=0.5pt] (0.2,-2) --(2.8,-2);
\draw[color=black!40,fill] (1,-2) circle (.25);
\node at (0.98,-2) {\footnotesize $+2$};
\draw[color=blue!40,fill] (2,-2) circle (.25);
\node at (1.98,-2) {\footnotesize $-1$};
\draw[dashed,color=gray,line width=0.5pt] (4.2,-2) --(6.8,-2);
\draw[color=blue!40,fill] (5,-2) circle (.25);
\node at (4.98,-2) {\footnotesize $+1$};
\draw[color=black!40,fill] (6,-2) circle (.25);
\node at (5.98,-2) {\footnotesize $-2$};
\draw[<->] (1.5,0.3) arc (135: 45 : 2.8);
\coordinate[label=below:$1$] (1) at (3.5,0.95);
\draw[<->] (1.5,-2.3) arc (-135: -45 : 2.8);
\end{tikzpicture}
\quad
\begin{tikzpicture}
\draw[dashed,color=gray,line width=0.5pt] (0.2,0) --(2.8,0);
\draw[color=black!40,fill] (1,0) circle (.25);
\node at (0.98,0) {\footnotesize $-2$};
\draw[color=black!40,fill] (2,0) circle (.25);
\node at (1.98,0) {\footnotesize $-2$};
\draw[dashed,color=gray,line width=0.5pt] (4.2,0) --(6.8,0);
\draw[color=blue!40,fill] (5,0) circle (.25);
\node at (4.98,0) {\footnotesize $+1$};
\draw[color=blue!40,fill] (6,0) circle (.25);
\node at (5.98,0) {\footnotesize $+1$};
\draw[dashed,color=gray,line width=0.5pt] (0.2,-2) --(2.8,-2);
\draw[color=blue!40,fill] (1,-2) circle (.25);
\node at (0.98,-2) {\footnotesize $-1$};
\draw[color=blue!40,fill] (2,-2) circle (.25);
\node at (1.98,-2) {\footnotesize $-1$};
\draw[dashed,color=gray,line width=0.5pt] (4.2,-2) --(6.8,-2);
\draw[color=black!40,fill] (5,-2) circle (.25);
\node at (4.98,-2) {\footnotesize $+2$};
\draw[color=black!40,fill] (6,-2) circle (.25);
\node at (5.98,-2) {\footnotesize $+2$};
\draw[<->] (1.5,0.3) arc (135: 45 : 2.8);
\coordinate[label=below:$1$] (1) at (3.5,0.95);
\draw[<->] (1.5,-2.3) arc (-135: -45 : 2.8);
\coordinate[label=below:$1$] (2) at (3.5,-2.5);
\draw[<->] (1.5,-0.3) arc (150: 210 : 1.45);
\coordinate[label=left:$1$] (1) at (1.2,-1.0);
\draw[<->] (5.5,-0.3) arc (30: -30 : 1.45);
\coordinate[label=right:$1$] (1) at (5.8,-1.0);
\end{tikzpicture}}
\caption{Transition rates of the particles on sites $N$ and $1$}\label{Fig:bulk:twist}
\end{figure}
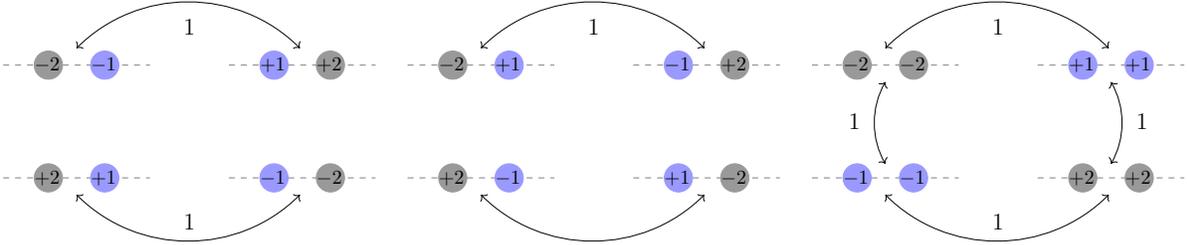

It should be remarked that particles on sites $N$ and $1$ exhibit transition rates that differ significantly from those in the bulk. When particles on these two sites have different colors, they may undergo a transformation into a new particle pair carrying the opposite charge relative to the original configuration. When particles occupying these two sites share identical configurations (colors and charges), they may transition into a new particle pair that maintains the same configuration but exhibits different colors relative to the original pair.
The transition rates on sites $N$ and $1$ are detailed in Figure \ref{Fig:bulk:twist}.

\subsection{Integrability of the model}
We can construct the following transfer matrix \cite{Yung95}
\bea
\bar t(u)={\rm tr}_0\{\mc{G}_0T_0(u)\},\label{tut}
\eea
where $T_0(u)$ is defined in \eqref{T1} and $\mc{G}$ reads
\begin{align}
&\mc{G}= 
\sigma_x \otimes \sigma_x. \label{tbm}
\end{align}
With the help of YBE \eqref{YBE} and the identity
\bea [R(u),\,\,\mc{G}\otimes \mc{G}]=0,\label{RG}\eea
it can be proven that the transfer matrices with different spectral parameters commute with each other, i.e.,
$[\,\bar t(u),\,\bar t(v)]=0.$ 
The transition matrix $\mc{\overline M}$ in \eqref{mt} can be obtained from the transfer matrix $\bar t(u)$
\begin{align}
\mc{\overline M}&=\left.\frac{\partial \ln \bar t(u)}{\partial
u}\right|_{u=0} -2N\times \mathbb{I}.
 \label{hhmt}
\end{align}

\subsection{Exact solutions}
\subsubsection{Decomposition of the transfer matrix}

Similar to the periodic case, the transfer matrix in Eq.~\eqref{tut} admits a factorization into the product of two sub-transfer matrices
\bea \bar t(u)=\bar t_{\sigma}(u)\otimes \bar{t}_{\tau}(u)\,, \label{apot} \eea
where  $\bar t_{s}(u)$ are defined as
\begin{equation}
\bar t_{s}(u)= {\rm tr}_{\bar 0} \{ \sigma^x_{\bar 0}\,\mc{T}^{s}_{\bar 0} (u)\}. \label{ts-1t}
\end{equation}
By using the Yang-Baxter relation \eqref{ai2-1p} and the identity
\bea  [\mc{R}^s(u),\,\,\sigma_x\otimes \sigma_x]=0,\label{Rg}   \eea 
 we can prove 
\bea
[\,\bar t_{s}(u),\, \bar t_{s'}(v)]=0, \quad s,s'=\sigma,\tau.
\eea
The derivative of $\ln \bar t_{\sigma,\tau}(u)$ gives the transition matrix of SSEP with twisted boundary \cite{Batchelor95,Yung95}
\begin{align}
\overline {\mc M}_{\sigma}&=\left.\frac{\partial \ln \bar t_{\sigma}(u)}{\partial
u}\right|_{u=0} -N\times \mathbb{I}\no\\
&=\frac12\sum_{j=1}^{N-1}(\sigma^x_j\sigma^x_{j+1}+\sigma^y_j\sigma^y_{j+1}+
\sigma^z_j\sigma^z_{j+1}-\mathbb{I})\no\\
&\quad +\frac12(\sigma^x_N\sigma^x_{1}-\sigma^y_N\sigma^y_{1}-
\sigma^z_N\sigma^z_{1}-\mathbb{I}),\\
\overline{\mc{M}}_{\tau}&=\overline{\mc{M}}_{\sigma}|_{
\{\sigma_j^\alpha\}\rightarrow \{\tau_j^\alpha,  \}}, \quad \alpha=x,y,z.\end{align} 
The transition matrix $\mc{\overline M}$ is the direct sum of $\overline {\mc M}_{\sigma}$ and $\overline {\mc M}_{\tau}$
\bea \mc{\overline M}=\overline{\mc{M}}_{\sigma}\oplus
\overline{\mc{M}}_{\tau}. \label{apop1t}
\eea
\begin{remark}
The twisted boundary condition in the SSEP can be understood as follows. In the bulk, hopping between adjacent sites is permitted only if they are in different states (one occupied, one empty). If the sites share the same state, hopping is forbidden. However, a different rule applies to the twisted boundary between sites $N$ and $1$: a transition occurs only if both sites are in the same state. Specifically, two particles can transition into two holes, and vice versa.
\end{remark}

\subsubsection{Exact solutions}

\paragraph{$T$-$Q$ relation}

The eigenvalue of
transfer matrix \eqref{tut} shares the same decomposition property \bea  \bar \Lambda(u)=
\bar \Lambda_\sigma(u) \bar\Lambda_{\tau}(u),\label{1BA012t}\eea where $\bar \Lambda_{\sigma}(u)$ and $\bar\Lambda_{\tau}(u)$ are the eigenvalues of transfer matrices $\bar t_\sigma(u)$ and $\bar t_\tau(u)$ in Eq. \eqref{ts-1t}, respectively. 

We can construct the following $T$-$Q$ relation for $\bar\Lambda_{\sigma}(u)$
\begin{align} \bar\Lambda_{\sigma}(u)=(u+1)^N\,\frac{ \bar Q(u-1)}{\bar Q(u)}-u^N\, \frac{ \bar Q(u+1)}{ \bar Q(u)}, \label{tsert}
\end{align}
where
\begin{align}
\bar Q(u)=\prod_{l=1}^{M}(u-\bar \mu_l),\quad 0\leq M\leq N.
\end{align}
The Bethe roots $\{\bar\mu_l\}$ in Eq. \eqref{tsert} should satisfy the BAEs
\bea
\left[\frac{\bar\mu_k}{\bar\mu_k+1}\right]^N=-
 \prod_{l\ne k}^{M}\frac{(\bar\mu_k-\bar\mu_l-1)}
 {(\bar\mu_k-\bar\mu_l+1)},\,
 \qquad k=1,\ldots,M.\label{BA012t}
 \eea

The eigenvalue of ${\bar t}_{\tau}(u)$, namely $\bar\Lambda_{\tau}(u)$, can be obtained directly from $\bar\Lambda_{\sigma}(u)$ with the following substitution 
\bea
\{ \bar\mu_j,M\}\rightarrow \{\bar \nu_j,{M}'\}.\eea
The eigenvalue of
transition matrix  \eqref{mt} in terms of the Bethe roots is
\begin{align} \bar E=&\left.\frac{\partial \ln (\Lambda_{\sigma}(u)\Lambda_{\tau}(u))}{\partial u}\right|_{u=0}-2N\no\\
=&\sum_{k=1}^{M}\frac{1}{\bar\mu_k(\bar\mu_k+1)}+\sum_{k=1}^{M'}\frac{1}{\bar\nu_k(\bar\nu_k+1)}.\label{energyt}
\end{align}

The independence of $\bar t_\tau(u)$ and $\bar t_\sigma(u)$ allows us to readily obtain all $2^N \times 2^N\equiv 4^N$ eigenvalues of $\bar t(u)$. Numerical solutions of BAEs in \eqref{BA012t} are shown in Table \ref{Tab2}. The resulting eigenvalues from \eqref{energyt} agree well with those obtained by exact diagonalization of the transition matrix.

\paragraph{Bethe states} The eigenstate of $\bar{t}(u)$, is also factorized and admits a parameterization as the following Bethe state
\begin{align}
|\bar\mu_1,\ldots,\bar\mu_M\rangle_\sigma\otimes |\bar\nu_1,\ldots,\bar\nu_{M'}\rangle_\tau,\\
\langle\bar\mu_1,\ldots,\bar\mu_M|_\sigma\otimes \langle\bar\nu_1,\ldots,\bar\nu_{M'}|_\tau,
\end{align}
where
\begin{align}
|u_1,\ldots,u_m\rangle_s&=\prod_{k=1}^n\bar{ \mathcal{B}}_s(u_k)\binom{1}{1}^{\otimes N},\quad 
\mbox{or}\quad |u_1,\ldots,u_m\rangle_s=\prod_{k=1}^n\bar{ \mathcal{C}}_s(u_k)\binom{1}{-1}^{\otimes N},\\
\langle u_1,\ldots,u_m|_s&=(1,\,1)^{\otimes N}\,\prod_{k=1}^m\bar{\mathcal{C}}_s(u_k),\quad
\mbox{or}\quad \langle u_1,\ldots,u_m|_s=(1,\,-1)^{\otimes N}\prod_{k=1}^m\bar{\mathcal{B}}_s(u_k),
\end{align}
and $\bar{\mathcal{B}}_s(u)$ and $\bar{\mathcal{C}}_s(u)$ are defined as
\begin{align}
\bar{\mathcal{B}}_s(u)&= \mathcal{A}_s(u)-\mathcal{B}_s(u)+\mathcal{C}_s(u)-\mathcal{D}_s(u),\\
\bar{\mathcal{C}}_s(u)&= \mathcal{A}_s(u)+\mathcal{B}_s(u)-\mathcal{C}_s(u)-\mathcal{D}_s(u).
\end{align}

\paragraph{Steady state}
The twisted boundary breaks the ${SU}(2)$ symmetry of the system described by $\overline{\mathcal{M}}_{s}$, while a ${Z}_2$ symmetry remains. The empty rapidity sets
$$
\{\bar\mu_1,\ldots,\bar\mu_M\}=\emptyset,\quad \{\bar\nu_1,\ldots,\bar\nu_{M'}\}=\emptyset,$$
now results in four null eigenstates of  $\overline{\mathcal{M}}$
\begin{align}
|\Psi_1\rangle=\frac{1}{4^N}\left(
\begin{array}{c}
1 \\[-2pt]
1 \\[-2pt]
1 \\[-2pt]
1 \\[-2pt]
\end{array}
\right)^{\otimes N},\qquad |\Psi_2\rangle=\frac{1}{4^N}\left(
\begin{array}{c}
1 \\[-2pt]
-1 \\[-2pt]
1 \\[-2pt]
-1 \\[-2pt]
\end{array}
\right)^{\otimes N},\\
|\Psi_3\rangle=\frac{1}{4^N}\left(
\begin{array}{c}
1 \\[-2pt]
1 \\[-2pt]
-1 \\[-2pt]
-1 \\[-2pt]
\end{array}
\right)^{\otimes N},\quad |\Psi_4\rangle=\frac{1}{4^N}\left(
\begin{array}{c}
1 \\[-2pt]
-1 \\[-2pt]
-1 \\[-2pt]
1 \\[-2pt]
\end{array}
\right)^{\otimes N}.
\end{align}
 
For a given physical initial state, the system evolves into a specific steady state in the long-time limit \( t \to \infty \), 
which must be a linear combination of the states \(\{|\Psi_k\rangle|k=1,2,3,4\}\)
\begin{align}
\lim_{t\to\infty}\ee^{\overline{\mathcal M}\,t}|\Phi\rangle&=\sum_{k=1}^4\frac{|\Psi_k\rangle\langle\Psi_k|\Phi\rangle}{\langle\Psi_k|\Psi_k\rangle}=4^N\sum_{k=1}^4\langle\Psi_k|\Phi\rangle|\Psi_k\rangle\no\\
&=\sum_{j_1,\ldots,j_N}w_{j_1,\ldots,j_N}(t)\bigotimes_{k=1}^N|j_k\rangle,\quad w_{j_1,\ldots,j_N}(t)\geq 0.
\end{align}
We see that the steady state to which the system evolves is completely determined by the initial condition.
As an example, we focus on the $N=4$ case and consider the long-time evolution starting from the following initial states
\begin{align}
\begin{aligned}
&|\Phi\rangle=|-2\rangle^{\otimes N}:\qquad \lim_{t\to \infty}\ee^{\overline{\mathcal M}\,t}|\Phi\rangle=|\Psi_1\rangle+|\Psi_2\rangle+|\Psi_3\rangle+|\Psi_4\rangle,\\
&|\Phi\rangle=|-1\rangle^{\otimes N}:\qquad \lim_{t\to \infty}\ee^{\overline{\mathcal M}\,t}|\Phi\rangle=|\Psi_1\rangle+|\Psi_2\rangle+|\Psi_3\rangle+|\Psi_4\rangle,\\
&|\Phi\rangle=\left(\tfrac{|-2\rangle+|-1\rangle}{2}\right)^{\otimes N}:\qquad \lim_{t\to \infty}\ee^{\overline{\mathcal M}\,t}|\Phi\rangle=|\Psi_1\rangle+|\Psi_3\rangle,\\
&|\Phi\rangle=\left(\tfrac{|-2\rangle+|-1\rangle+|+1\rangle}{3}\right)^{\otimes N}:\qquad \lim_{t\to \infty}\ee^{\overline{\mathcal M}\,t}|\Phi\rangle=\frac{1}{9}(9|\Psi_1\rangle+\frac19|\Psi_2\rangle+\frac19|\Psi_3\rangle+\frac19|\Psi_4\rangle),\\
&|\Phi\rangle=|-2\rangle\otimes|-1\rangle\otimes|-2\rangle\otimes|+1\rangle:\quad \lim_{t\to \infty}\ee^{\overline{\mathcal M}\,t}|\Phi\rangle=|\Psi_1\rangle-|\Psi_2\rangle-|\Psi_3\rangle+|\Psi_4\rangle,\\
&|\Phi\rangle=|-2\rangle\otimes|-1\rangle\otimes|-2\rangle\otimes|-2\rangle:\quad \lim_{t\to \infty}\ee^{\overline{\mathcal M}\,t}|\Phi\rangle=|\Psi_1\rangle-|\Psi_2\rangle+|\Psi_3\rangle-|\Psi_4\rangle.
\end{aligned}
\end{align}
Since $\bar t(u)$ (and $\overline{\mathcal{M}}$) is Hermitian, we can easily get the left steady states.  
For any steady state \(|\Psi\rangle = \sum_{k=1}^4 c_k |\Psi_k\rangle\) (\(c_k \in \mathbb{R}\)), the \(n\)-point correlation functions (with \(n < N\)) are given by the same expression
\begin{align}
\langle\hat{n}_{j_1,\a_1}\hat{n}_{j_2,\a_2}\cdots \hat{n}_{j_n,\a_n}\rangle&=\frac{\langle\Psi|\hat{n}_{j_1,\a_1}\hat{n}_{j_2,\a_2}\cdots \hat{n}_{j_n,\a_n}|\Psi\rangle}{\langle\Psi|\Psi\rangle}\no\\
&=\frac{\sum_{k',k=1}^4c_{k'}{c_k}\langle\Psi_{k'}|\hat{n}_{j_1,\a_1}\hat{n}_{j_2,\a_2}\cdots \hat{n}_{j_n,\a_n}|\Psi_k\rangle}{\sum_{k',k=1}^4c_{k'}{c_k}\langle\Psi_{k'}|\Psi_k\rangle}\no\\
&=\frac{\sum_{k=1}^4{c^2_k}\langle\Psi_{k}|\hat{n}_{j_1,\a_1}\hat{n}_{j_2,\a_2}\cdots \hat{n}_{j_n,\a_n}|\Psi_k\rangle}{\sum_{k=1}^4{c^2_k}\langle\Psi_{k}|\Psi_k\rangle}=\frac{1}{4^n},
\end{align}
where $j_l=1,\ldots,N,\,\,j_{l}\neq j_{l'},$ and $\a_l=\pm2,\pm1$.
\begin{table}[htbp]
    \centering
    \begin{tabular}{|c|c|}
    \hline
      $\bar\mu_1/\bar\nu_1$   &  $\bar\mu_2/\bar\nu_2$ \\
      \hline 
       --- & --- \\
$-$0.5000$-$1.2071i   & --- \\
$-$0.5000+1.2071i & --- \\
$-$0.5000$-$0.2071i & --- \\
$-$0.5000+0.2071i & --- \\
$-$0.5000$-$0.8660i & $-$0.5000+0.8660i \\
 $-$0.5000$-$0.0841i & $-$0.5000+0.7021i \\
 $-$0.5000+0.0841i & $-$0.5000$-$0.7021i \\
 $-$1.1360+0.8090i & ~~0.1360+0.8090i \\
 $-$1.1360$-$0.8090i & ~~0.1360$-$0.8090i \\
 $-$1.0000 & 0.0000 \\
    \hline
    \end{tabular}
    \caption{Numeric solutions of BAEs \eqref{BA012t} with $N=4$.}
    \label{Tab2}
\end{table}

\section{Integrable symmetric stochastic process with open boundary conditions}\label{sec:open}
\setcounter{equation}{0}

\subsection{Transition matrix}
We now consider a stochastic process with open boundaries, where particles can enter and exit the system at both boundaries.  
The evolution of the model is governed by the following master equation. 
\bea\frac{d}{dt}|\Psi(t)\rangle =\widetilde{\mc{M}}|\Psi(t)\rangle.\label{Master:Open} \eea
The transition matrix $\mc{\widetilde M}$ in (\ref{Master:Open}) is
\bea &&\mc{\widetilde M}=\mc{M}_1+\sum_{k=1}^{N-1}\mc{M}_{k,k+1}+\mc{M}_N,\label{m}
\eea
where $\mathcal{M}_{k,k+1}$ is given in Eq. \eqref{Mkk1}, and $\mathcal{M}_1$ and $\mathcal{M}_N$ describe the transitions at the left and right boundaries, respectively, and read
\begin{align}
\mc{M}_1&= \left(\begin{array}{cccc}-s_1-t_1 &t_2&s_2 &0 \\
t_1& -s_1-t_2  &0 &s_2 \\
s_1 &0 &-s_2-t_1 &t_2\\
0 &s_1 &t_1&-s_2-t_2
\end{array}\right),\no\\
\mc{M}_N&= \left(\begin{array}{cccc}s'_1+t'_1 &-t'_2&-s'_2 &0 \\
-t'_1& s'_1+t'_2  &0 &-s'_2 \\
-s'_1 &0 &s'_2+t'_1 &-t'_2\\
0 &-s'_1 &-t'_1&s'_2+t'_2
\end{array}\right).\no
\end{align}

\begin{figure}[htbp]
\centering
\begin{tikzpicture}
\draw[dashed,color=gray,line width=0.5pt] (3,0) --(4.8,0);
\draw[dashed,color=gray,line width=0.5pt] (3,-0.5) --(3,0.5);
\draw[color=blue!40,fill] (1,0) circle (.25);
\node at (0.98,0) {\footnotesize $-1$};
\draw[color=blue!40,fill] (2,0) circle (.25);
\node at (1.98,0) {\footnotesize $+1$};
\draw[color=black!40,fill] (4,0) circle (.25);
\node at (3.98,0) {\footnotesize $-2$};
\draw[<->] (1.0,-0.3) arc (-160 : -20 : 1.6);
\coordinate[label=above:$t_1$] (1) at (2.5,-1.2);
\draw[<->] (2.0,0.3) arc (160 : 20 : 1.05);
\coordinate[label=below:$s_1$] (1) at (3,1.0);
\end{tikzpicture}\hspace{3.0cm}
\begin{tikzpicture}
\draw[dashed,color=gray,line width=0.5pt] (3,0) --(4.8,0);
\draw[dashed,color=gray,line width=0.5pt] (3,-0.5) --(3,0.5);
\draw[color=black!40,fill] (1,0) circle (.25);
\node at (0.98,0) {\footnotesize $-2$};
\draw[color=black!40,fill] (2,0) circle (.25);
\node at (1.98,0) {\footnotesize $+2$};
\draw[color=blue!40,fill] (4,0) circle (.25);
\node at (3.98,0) {\footnotesize $-1$};
\draw[<->] (1.0,-0.3) arc (-160 : -20 : 1.6);
\coordinate[label=above:$t_2$] (1) at (2.5,-1.2);
\draw[<->] (2.0,0.3) arc (160 : 20 : 1.05);
\coordinate[label=below:$s_1$] (1) at (3,1.0);
\end{tikzpicture}\hspace{3.0cm}
\begin{tikzpicture}
\draw[dashed,color=gray,line width=0.5pt] (3,0) --(4.8,0);
\draw[dashed,color=gray,line width=0.5pt] (3,-0.5) --(3,0.5);
\draw[color=black!40,fill] (1,0) circle (.25);
\node at (0.98,0) {\footnotesize$-2$};
\draw[color=black!40,fill] (2,0) circle (.25);
\node at (1.98,0) {\footnotesize$+2$};
\draw[color=blue!40,fill] (4,0) circle (.25);
\node at (3.98,0) {\footnotesize $+1$};
\draw[<->] (1.0,-0.3) arc (-160 : -20 : 1.6);
\coordinate[label=above:$s_2$] (1) at (2.5,-1.2);
\draw[<->] (2.0,0.3) arc (160 : 20 : 1.05);
\coordinate[label=below:$t_1$] (1) at (3,1.0);
\end{tikzpicture}\hspace{3.0cm}
\begin{tikzpicture}
\draw[dashed,color=gray,line width=0.5pt] (3,0) --(4.8,0);
\draw[dashed,color=gray,line width=0.5pt] (3,-0.5) --(3,0.5);
\draw[color=blue!40,fill] (1,0) circle (.25);
\node at (0.98,0) {\footnotesize$-1$};
\draw[color=blue!40,fill] (2,0) circle (.25);
\node at (1.98,0) {\footnotesize$+1$};
\draw[color=black!40,fill] (4,0) circle (.25);
\node at (3.98,0) {\footnotesize$+2$};
\draw[<->] (1.0,-0.3) arc (-160 : -20 : 1.6);
\coordinate[label=above:$s_2$] (1) at (2.5,-1.2);
\draw[<->] (2.0,0.3) arc (160 : 20 : 1.05);
\coordinate[label=below:$t_2$] (1) at (3,1.0);
\end{tikzpicture}
\caption{Transition rates of the particles on the left boundary. Particles on the right boundary exhibit analogous behavior, with corresponding transition rates obtained through the parameter substitutions $t_i\to -t'_i$, $s_i\to -s'_i$. }\label{Fig:left}
\end{figure}
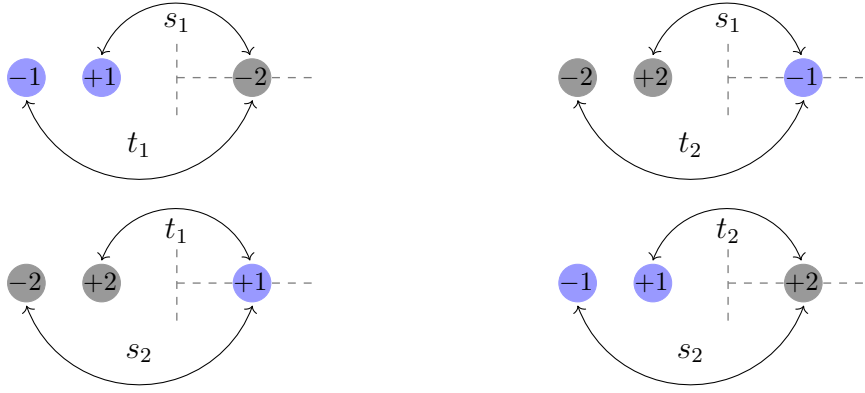

At the boundaries, particles can enter and exit the system. Once a site is occupied by a particle of a particular color, it cannot be replaced by another particle of the same color. More details about the transition rates on the boundaries are presented in Figure \ref{Fig:left}. 
To ensure all the transition rates are non-negative, we assume $t_1,t_2,s_1,s_2\geq 0$ and $t'_1,t'_2,s'_1,s'_2\leq 0$.

\subsection{Integrability of the model}

Let us first introduce the $K$-matrix on the left boundary 
\bea
K^{-}(u)=\left(\begin{array}{cccc}k_{11}(u) &k_{12}(u)&k_{13}(u) &k_{14}(u) \\
k_{21}(u) &k_{22}(u)&k_{23}(u) &k_{24}(u) \\
k_{31}(u) &k_{32}(u) &k_{33}(u) &k_{34}(u)\\
k_{41}(u) &k_{42}(u)&k_{43}(u) &k_{44}(u)
\end{array}\right),\label{K-matrix-12}\eea
which satisfies the reflection equation (RE) \cite{Skl88,Cherednik84}
\begin{equation}
 R _{1,2}(u-v){K^{-}_{  1}}(u)R _{2,1}(u+v) {K^{-}_{2}}(v)=
 {K^{-}_{2}}(v)R _{1,2}(u+v){K^{-}_{1}}(u)R _{2,1}(u-v).
 \label{r1}
 \end{equation}
The elements of $K^-(u)$ in Eq. \eqref{K-matrix-12} are
\bea&& k_{11}(u)=(1-(s_1-s_2)u)(1-(t_1-t_2)u),\quad k_{12}(u)=2t_2u(1-(s_1-s_2)u),\no\\
&& k_{13}(u)=2s_2u(1-(t_1-t_2)u),\quad k_{14}(u)=4s_2t_2u^2,\no\\
&&k_{21}(u)=2t_1u(1-(s_1-s_2)u),\quad k_{22}(u)=(1-(s_1-s_2)u)(1+(t_1-t_2)u),\no\\
&&k_{23}(u)=4s_2t_1u^2,\quad k_{24}(u)=2s_2u(1+(t_1-t_2)u),\no\\
&&k_{31}(u)=2s_1u(1-(t_1-t_2)u),\quad k_{32}(u)=4s_1t_2u^2,\no\\
&&k_{33}(u)=(1+(s_1-s_2)u)(1-(t_1-t_2)u),\quad k_{34}(u)=2t_2u(1+(s_1-s_2)u),\no\\
&&k_{41}(u)=4s_1t_1u^2,\quad k_{42}(u)=2s_1(1+(t_1-t_2)u),\no\\
&&k_{43}(u)=2t_1u(1+(s_1-s_2)u),\quad k_{44}(u)=(1+(s_1-s_2)u)(1+(t_1-t_2)u).\no
\eea

\begin{remark}
The $K$-matrix in Eq. \eqref{K-matrix-12} is not the most general solution to the reflection equation \eqref{r1}. In this work, we restrict our attention to the case where the associated transfer matrix yields a model that satisfies stochasticity.
\end{remark}

The boundary $K$-matrix on the right boundary can be obtained via the following
simple substitution 
\begin{equation}
K^{ +}(u)=K^{ -}(-u-1)|_{\{s_i,t_i\}\rightarrow
\{s'_i,t'_i\}}.\label{ksk12}
\end{equation} 
The matrix $K^+(u)$ satisfies the dual RE
\begin{eqnarray}
&R _{1,2}(-u+v){K^{ +}_{1}}(u)R _{2,1}
 (-u-v-2){K^{ +}_{2}}(v)\nonumber\\
&\qquad\qquad={K^{ +}_{2}}(v)R _{1,2}(-u-v-2) {K^{ +}_{1}}(u)R
_{2,1}(-u+v).\label{r2}
\end{eqnarray}
We can construct the transfer matrix $\tilde t(u)$ as \cite{Skl88}
\begin{equation}
\tilde t(u)={\rm tr}_0 \{K_0^{ +}(u)T_0 (u) K^{ -}_0(u)\widehat{T}_0
(u)\},\label{tru}
\end{equation}
where the one-row monodromy matrices are
\begin{align}
T_0 (u)&=R _{0,N}(u)R _{0,N-1}(u)\cdots R
_{0,1}(u),\label{T2}\\
\widehat{T}_0 (u)&=R_{1,0}(u)R_{2,0}(u)\cdots R_{N,0}(u).\label{Tt11}
\end{align}
The YBE \eqref{YBE} and the REs \eqref{r1}, \eqref{r2} together guarantee the mutual commutativity of the transfer matrices. Therefore, $\tilde t(u)$ serves as the generating function of
all the conserved quantities in the model.
The transition matrix $\mc{\widetilde M}$ in \eqref{m} can be obtained from the transfer matrix as follows
\begin{align}
\mc{\widetilde M}&=\frac{1}{2}\,\left.\frac{\partial \ln \tilde t(u)}{\partial
u}\right|_{u=0} -{\rm const}\times\mathbb{I}. \label{hhm}
\end{align}

\subsection{Exact solutions of the model}
\subsubsection{Decomposition of the transfer matrix}

The reflection matrices presented in Eqs. \eqref{K-matrix-12} and \eqref{ksk12} can also be expressed in a factorized form
\begin{align}
K^\pm(u)= \mc{K}^{\pm}(\sigma,u) \otimes  \mc{K}^{\pm}(\tau,u),\label{ktk}
\end{align}
where $\mc{K}^{\pm}(\sigma,u)$ and $\mc{K}^{\pm}(\tau,u)$ read \cite{Crampe14}
\begin{align} &\mc{K}^{-}(\sigma,u)=\left(\begin{array}{cc}
    1-(s_1-s_2)u&2s_2 u \\
   2s_1 u &1+(s_1-s_2)u
   \end{array}\right), \\[4pt]
&\mc{K}^{+}(\sigma,u)= \mc{K}^{-}(\sigma,-u-1)|_{\{s_1,s_2\}\rightarrow\{s'_1,s'_2\}}, \\[4pt]
&\mc{K}^{-}(\tau,u)=\left(\begin{array}{cc}1-(t_1-t_2)u&2t_2 u \\
2t_1 u &1+(t_1-t_2)u
   \end{array}\right), \\[4pt]
&\mc{K}^{+}(\tau,u)= \mc{K}^{-}(\tau,-u-1)|_{\{t_1,t_2\}\rightarrow\{t'_1,t'_2\}}. 
\end{align}
The $\mc{K}^{\pm}$-matrices satisfy the (dual) reflection relations
\begin{align}
 &\mc{R}^s _{1,2}(u-v){\mc{K}^{-}_{  1}}(s,u)\mc{R}^s_{2,1}(u+v) {\mc{K}^{-}_{2}}(s,v)\no\\
&={\mc{K}^{-}_{2}}(s,v)\mc{R}^s _{1,2}(u+v){\mc{K}^{-}_{1}}(s,u)\mc{R}^s _{2,1}(u-v),\label{r1s}\\
&\mc{R}^s_{1,2}(-u+v){\mc{K}^{ +}_{1}}(s,u)\mc{R}^s _{2,1}
 (-u-v-2){\mc{K}^{+}_{2}}(s,v)\nonumber\\
&={\mc{K}^{+}_{2}}(s,v)\mc{R}^s _{1,2}(-u-v-2) {\mc{K}^{+}_{1}}(s,u)\mc{R}^s
_{2,1}(-u+v).
 \label{r2s}
 \end{align}

It follows from Eqs. \eqref{rtr} and \eqref{ktk} that $\tilde{t}(u)$ admits a decomposition into the product of two mutually commutative transfer matrices, each corresponding to an XXX spin chain with open boundaries.
\bea \tilde t(u)=\tilde t_{\sigma}(u)\otimes \tilde{t}_{\tau}(u)\,, \label{apo} \eea
where  $\tilde t_{s}(u)$ are defined as
\begin{align}
\tilde t_{s}(u)&= {\rm tr}_{\bar 0} \{ \mc{K}^{+}_{\bar 0}(s,u)\mathbb{T}_{\bar 0}^s(u)\},\label{ts-1}\\
\mathbb{T}_{\bar 0}^s(u)&=\mc{R}^s_{\bar 0,\bar N}(u)\cdots \mc{R}^s_{\bar 0,\bar 2}(u)\mc{R}^s_{\bar 0,\bar 1}(u)\mc{K}^{-}_{\bar 0}(s,u)\no\\
&\quad\times\mc{R}^s_{\bar 1,\bar 0}(u)\mc{R}^s_{\bar 2, 0'}(u)\cdots \mc{R}^s_{\bar N,\bar 1}(u)\no\\
&= \left(
\begin{array}{cc}
\mathbb{A}_s(u) & \mathbb{B}_s(u) \\
\mathbb{C}_s(u) & \mathbb{D}_s(u) 
\end{array}
\right).\label{MatrixForm:T}
\end{align}
The logarithm derivative of $\tilde t_{\sigma,\tau}(u)$ gives the transition matrix of SSEP under generic open boundary condition
\begin{align}
\widetilde{\mc {M}}_\sigma&=\frac{1}{2}\,\left.\frac{\partial \ln \tilde t_\sigma(u)}{\partial
u}\right|_{u=0} -\rm{const}\times \mathbb{I}\no\\
&=\frac12\sum_{j=1}^{N-1}(\sigma^x_j\sigma^x_{j+1}+\sigma^y_j\sigma^y_{j+1}+
\sigma^z_j\sigma^z_{j+1}-\mathbb{I})\no\\
&\quad +s_2\sigma_1^++s_1\sigma_1^- +\frac{1}{2} (s_2-s_1)\sigma^z_1-\frac{1}{2}(s_1+s_2)\mathbb{I}\no\\
&\quad -s'_2\sigma_N^+-s'_1\sigma_N^- -\frac{1}{2} (s'_2-s'_1)\sigma^z_N+\frac{1}{2}(s'_1+s'_2)\mathbb{I},\,\\
\widetilde{\mc{M}}_{\tau}&=\widetilde{\mc M}_{\sigma}|_{
\{\sigma_j^\alpha, s_i, s'_i\}\rightarrow \{\tau_j^\alpha,  t_i,
t'_i\}}, \quad \alpha=x,y,z,\quad i=1,2.\end{align} 
Hence, the transition matrix in Eq. \eqref{m} is the direct sum of the transition matrices of two independent open SSEPs
 \bea \mc{\widetilde M}=\widetilde{\mc M}_{\sigma}\oplus
\widetilde{\mc M}_{\tau}. \label{apop1}\eea

\subsubsection{Exact solutions}
\paragraph{$T$-$Q$ relation}
The decomposition of the transfer matrix \eqref{tru} implies that its eigenvalue processes the similar property \bea  \tilde \Lambda(u)=
\tilde\Lambda_{\sigma}(u) \tilde\Lambda_{\tau}(u),\label{1BA012}\eea where $\tilde\Lambda_{s}(u)$ is the
 eigenvalue of the 
transfer matrix $\tilde t_s(u)$ in Eq.  \eqref{ts-1}.
We can construct the following $T$-$Q$ relation for $\tilde\Lambda_{\sigma}(u)$ \cite{Belliard13}
\begin{align} \tilde\Lambda_{\sigma}^{\pm}(u)=&\,\frac{2u+2}
{2u+1}(1\pm w_1u)(1\mp w_2u)(u+1)^{2N}\frac{ \widetilde Q_\pm(u-1)}{ \widetilde Q_\pm(u)}\no\\[4pt]
&\,+\frac{2u}
{2u+1}[1\mp w_1(u+1)][1\pm w_2(u+1)]u^{2N} \frac{\widetilde Q_\pm(u+1)}{\widetilde Q_\pm(u)}, \label{tser}
\end{align}
where
\bea
&& w_1=s_1+s_2,\quad w_2=s'_1+s'_2,\\
&& \widetilde Q_\pm(u)=\prod_{l=1}^{M_\pm}(u-\tilde\mu^{\pm}_l)(u+\tilde\mu^{\pm}_l+1).
\eea
The Bethe roots $\{\tilde\mu^\pm_l\}$ in Eq. \eqref{tser} are determined by the Bethe ansatz equations
\bea
&&\frac{[1\mp w_1(\tilde\mu_k^{\pm}+1)][1\pm w_2(\tilde\mu_k^{\pm}+1)]}{[1\pm w_1\tilde\mu_k^{\pm}][1\mp w_2\tilde\mu_k^{\pm}]}\left[\frac{\tilde\mu_{k}^{\pm}}{\tilde\mu_{k}^{\pm}+1}\right]^{2N}\no\\
 &&=
 \prod_{l\ne k}^{M_\pm}\frac{(\tilde\mu^{\pm}_k-\tilde\mu^{\pm}_l-1)(\tilde\mu^{\pm}_k+\tilde\mu^{\pm}_l)}
 {(\tilde\mu^{\pm}_k-\tilde\mu^{\pm}_l+1)(\tilde\mu^{\pm}_k+\tilde\mu^{\pm}_l+2)},\,
 \qquad k=1,\cdots, M_\pm. \label{BA012}\eea
Here, $\tilde{\Lambda}_\sigma^+(u)$ and $\tilde{\Lambda}_\sigma^-(u)$ are equivalent, and they represent different parameterizations of the same polynomial $\tilde{\Lambda}_\sigma(u)$. The integer $M_\pm$ in the $T$-$Q$ relation can take values from 0 to $N$.

The eigenvalue of $\tilde t_\tau(u)$, namely $\tilde\Lambda_\tau^\pm(u)$ can be obtained directly from $\tilde\Lambda_\sigma^\pm(u)$ with the following substitution 
\bea
\{s_1, s_2,  s'_1, s'_2,\tilde\mu_j^\pm,M_\pm\}\rightarrow \{t_1, t_2,  t'_1, t'_2,\tilde\nu_j^\pm,M'_\pm\}.\label{Open:Substitution}\eea
Then the  exact eigenvalue of
transition matrix  \eqref{m} can be obtained
\begin{align} \tilde E_{\epsilon_1,\epsilon_2}=&\,\frac12\left[\frac{\partial \ln (\tilde\Lambda^{\epsilon_1}_{\sigma}(u)\tilde\Lambda^{\epsilon_2}_{\tau}(u))}{\partial u}\right|_{u=0}-4N+2\no\\
&\,-(s_1+s_2+t_1+t_2)+(s'_1+s'_2+t'_1+t'_2)]\no\\
=&\sum_{k=1}^{M_{\epsilon_1}}\frac{1}{\tilde\mu_k^{\epsilon_1}(\tilde\mu_k^{\epsilon_1}+1)}+\sum_{k=1}^{ M'_{\epsilon_2}}\frac{1}{\tilde\nu_k^{\epsilon_2}(\tilde\nu_k^{\epsilon_2}+1)}\no\\
&+\frac{\epsilon_1}{2}(s_1+s_2-s'_1-s'_2)+\frac{\epsilon_2}{2}(t_1+t_2-t'_1-t'_2)\no\\
&-\frac12(s_1+s_2+t_1+t_2)+\frac12(s'_1+s'_2+t'_1+t'_2),\label{energy}
\end{align}
where $\epsilon_1,\epsilon_2=\pm$. The numerical solutions of BAEs \eqref{BA012} are shown in Table \ref{Tab3}.
\begin{table}[htbp]
\centering
\begin{tabular}{|c|c|c|}
\hline
  $\tilde\mu_1^+$& $\tilde\mu_2^+$ & $\tilde\mu_3^+$ \\
\hline 
-- & -- & --\\
$-$0.5000$-$1.3185i & -- & -- \\
$-$0.5000$-$0.2299i & -- & -- \\
$-$0.5000$-$0.5417i & -- & -- \\
0.0257+0.8645i & 0.0257$-$0.8645i & --\\
 0.0000+0.2945i & 0.0000$-$0.2945i & --\\
 $-$0.5000+0.2488i & $-$0.5000+0.7455i & --\\
\hline 0.0000+0.2548i & 0.0000$-$0.2548i & 1.8004  \\
\hline
\end{tabular}

\begin{tabular}{|c|c|c|}
\hline
  $\tilde\mu_1^-$& $\tilde\mu_2^-$ & $\tilde\mu_3^-$ \\
\hline 
-- & -- & --\\
$-$0.5000$-$2.4379i & -- & --\\
$-$0.5000$-$0.3337i & -- & --\\
0.8758 & -- & -- \\
0.0061$-$0.3777i& 0.0061 + 0.3777i & --\\ $-$0.5000+0.8437i & 0.6777 & -- \\
0.8129+0.2412i & 0.8129$-$0.2412i & --\\
0.6523+0.4982i & 0.6523$-$0.4982i & 0.6926 \\
\hline
\end{tabular}

\begin{tabular}{|c|c|c|}
\hline
  $\tilde\nu_1^+$& $\tilde\nu_2^+$ & $\tilde\nu_3^+$ \\
\hline 
-- & -- & --\\
$-$0.5000$-$1.0006i& -- & --\\
$-$0.5000$-$0.3946i &-- & --\\
$-$0.5000$-$0.1510i &-- & --\\
$-$0.5000$-$0.1664i & $-$0.5000+0.5070i & --\\
0.0002$-$0.3080i & 0.0002+0.3080i &  --\\
0.4431 & $-$0.4576 &  --\\
$-$0.5000+0.0842i &$-$0.5000+1.7186i & 0.5180 \\
\hline
\end{tabular}
\begin{tabular}{|c|c|c|}
\hline
  $\tilde\nu_1^-$& $\tilde\nu_2^-$ & $\tilde\nu_3^-$ \\
\hline 
-- & -- & --\\
1.0494  & & \\
$-$0.5000+0.3582i & &\\
0.5734 &  & \\
0.1395+0.7199i & 0.1395$-$0.7199i &  \\
0.4961 & $-$0.5000+0.9399i & \\
0.5554$-$0.1324i & 0.5554+0.1324i &  \\
0.4880$-$0.3351i &0.4880+0.3351i & 0.4758 \\
\hline
\end{tabular}
\caption{Numeric solutions of BAEs \eqref{BA012} with $N=3$, $\{s_1,s_2,t_1,t_2\}=\{0.36,0.52,0.66,0.81\}$ and  $\{s'_1,s'_2,t'_1,t'_2\}=\{-0.32,-0.48,-0.56,-0.90\}$.}\label{Tab3}
\end{table}

\paragraph{Bethe states}
Although both $\mc{K}^{+}(\sigma,u)$ and $\mc{K}^{-}(\sigma,u)$ are non-diagonal, the $T$-$Q$ relation retains the same form as in the diagonal case. This arises from the fact that $\mc{K}^+(\sigma,u)$ can be diagonalized and $\mc{K}^-(\sigma,u)$ can be triangularized simultaneously via the a proper gauge transformation
\begin{align}
 &g=\left(
\begin{array}{cc}
 s'_2 & -1 \\
 s'_1 & 1 \\
\end{array}
\right),\quad g^{-1}\mc{K}^-(\sigma,u)g= \left(
\begin{array}{cc}
 1+w_1u & 0 \\
 2u (s_1s'_2-s'_1s_2) & 1-w_1 u \\
\end{array}
\right),\no\\
&g^{-1}\mc{K}^+(\sigma,u)g=\left(
\begin{array}{cc}
 1-w_2(u+1) & 0 \\
 0 & 1+w_2(u+1)\\
\end{array}
\right),\\
 &\bar g=\left(
\begin{array}{cc}
 -1 & s'_2 \\
 1 & s'_1 \\
\end{array}
\right),\quad \bar g^{-1}\mc{K}^-(\sigma,u)\bar g= \left(
\begin{array}{cc}
 1-w_1u & 2u (s_1s'_2-s'_1s_2) \\
  0 & 1+w_1 u \\
\end{array}
\right),\no\\
&\bar g^{-1}\mc{K}^+(\sigma,u)\bar g=\left(
\begin{array}{cc}
 1+w_2(u+1) & 0 \\
 0 & 1-w_2(u+1)\\
\end{array}
\right).
\end{align}
The eigenstate of $\tilde t_\sigma(u)$  can be constructed via the algebraic Bethe ansatz method \cite{Korepin97,Belliard13}, specifically as 
\begin{align}
|\tilde\mu_1^-,\ldots,\tilde\mu_{M_-}^-\rangle_\sigma&=\widetilde{\mathbb{B}}_\sigma(\tilde\mu_1^-)\cdots\widetilde{\mathbb{B}}_\sigma(\tilde\mu_{M_-}^-)\binom{1}{-1}^{\otimes N},\label{Ket:BetheState}\\
\langle \tilde\mu_1^+,\ldots,\tilde\mu_{M_+}^+|_\sigma&=(1,\,1)^{\otimes N}\widetilde{\mathbb{B}}_\sigma(\tilde\mu_1^+)\cdots\widetilde{\mathbb{B}}_\sigma(\tilde\mu_{M_+}^+),\label{Bra:BetheState}
\end{align}
where 
\begin{align}
\widetilde{\mathbb{B}}_\sigma(u)= \mathbb{A}_\sigma(u) s'_1s'_2+\mathbb{B}_\sigma(u) s'_1s'_1-\mathbb{C}_\sigma(u) s'_2s'_2-\mathbb{D}_\sigma(u) s'_1s'_2,
\end{align}
and $\mathbb{A}(u)$, $\mathbb{B}(u)$, $\mathbb{C}(u)$ and $\mathbb{D}(u)$ are given by Eq. (\ref{MatrixForm:T}).

The eigenstate of $\tilde t(u)$ can be constructed as
\begin{align}  |\tilde\mu_1^-,\ldots,\tilde\mu^-_{M_-}\rangle_\sigma\otimes |\tilde\nu_1^-,\ldots,\tilde\nu^-_{M'_-}\rangle_\tau,\\
\langle\tilde\mu_1^+,\ldots,\tilde\mu^+_{M_+}|_\sigma\otimes |\tilde\nu_1^+,\ldots,\tilde\nu^+_{M'_+}\rangle_\tau,
\end{align}
where $|\tilde\nu_1^-,\ldots,\tilde\nu^-_{M'_-}\rangle_\tau$ and $|\tilde\nu_1^-,\ldots,\tilde\nu^-_{M'_-}\rangle_\tau$ can be obtained directly from Eqs. (\ref{Ket:BetheState}) and (\ref{Bra:BetheState}) with the substitution \eqref{Open:Substitution}.

\paragraph{Steady state} Under generic open boundary conditions, the steady state of the system is unique. The left steady state corresponds to the empty rapidity sets $$
\{\tilde\mu_1^+,\ldots,\tilde\mu^+_{M_+}\}=\emptyset,\quad \{\tilde\nu_1^+,\ldots,\tilde\nu^+_{M_-}\}=\emptyset,$$
and has the following factorized form 
\begin{align}
\langle \Psi|={ _\sigma}\langle\Psi|\otimes {_\tau}\langle\Psi|,\quad { _{s}}\langle\Psi|=(1,1)^{\otimes N},\quad s=\sigma,\tau.
\end{align}
Because $\tilde t(u)$ and $\widetilde{M}$ are non‑Hermitian, the right steady state $|\Psi\rangle$ is not the conjugate transpose of $\langle \Psi|$; it is considerably complex and can be expressed in the following form
\begin{align}
|\Psi\rangle=|\Psi\rangle_\sigma\otimes|\Psi\rangle_\tau,\quad 
|\Psi\rangle_\sigma=|\tilde\mu_1^-,\ldots,\tilde\mu_{N}^-\rangle_\sigma,\quad 
|\Psi\rangle_\tau=|\tilde\nu_1^-,\ldots,\tilde\nu_{N}^-\rangle_\tau,\label{Open:SteadyState}
\end{align}
where $\{\tilde\mu_1^-,\ldots,\tilde\mu_{N}^-\}$ and $\{\tilde\nu_1^-,\ldots,\tilde\nu_{N}^-\}$
are the Bethe roots corresponding to the steady state. The right steady state depends critically on the boundary parameters, and its properties have been studied using the Bethe ansatz and the matrix product ansatz methods \cite{Derrida98,Crampe14,Frassek_2020}.

The correlation functions in the steady state can also be derived by leveraging the known results for the open SSEP.
Let us define 
\begin{align}
\rho_a = \frac{s_1}{s_1 + s_2},
\quad 
\rho_{b} = \frac{s_1'}{s_1' + s_2'},
\quad 
\bar\rho_{a} = \frac{t_1}{t_1 + t_2},
\quad 
\bar\rho_{b} = \frac{t_1'}{t_1' + t_2'}.
\end{align}
The expectation value $\frac{{_s}\langle\Psi|\hat{n}_{k,\downarrow}|\Psi\rangle_s}{{_s}\langle\Psi|\Psi\rangle_s}$ can be expressed as the following formulas
\cite{Derrida02,Frassek_2020}
\begin{align}
\langle\hat{n}_{k,\downarrow}\rangle_\sigma&=\frac{\rho_{a} (N-k- \frac{1}{s_1' + s_2'})+ \rho_{b} (k-1 + \frac{1}{s_1 + s_2})}{N + \frac{1}{s_1 + s_2} - \frac{1}{s_1' + s_2'} - 1},\\
\langle\hat{n}_{k,\downarrow}\rangle_\tau&=\frac{\bar\rho_{a} (N-k- \frac{1}{t_1' + t_2'})+ \bar\rho_{b} (k-1 + \frac{1}{t_1 + t_2})}{N + \frac{1}{t_1 + t_2} - \frac{1}{t_1' + t_2'} - 1}.
\end{align}
Consequently, the system's density profile 
$\langle\hat{n}_{k,l}\rangle$ can be expressed directly as a product of the single-site density functions from each independent channel (or subsystem), specifically as follows
\begin{align}
\begin{aligned}
&\langle\hat{n}_{k,-2}\rangle=(1-\langle\hat{n}_{k,\downarrow}\rangle_\sigma)(1-\langle\hat{n}_{k,\downarrow}\rangle_\tau),\\
&\langle\hat{n}_{k,-1}\rangle=(1-\langle\hat{n}_{k,\downarrow}\rangle_\sigma)\langle\hat{n}_{k,\downarrow}\rangle_\tau,\\
&\langle\hat{n}_{k,+1}\rangle=\langle\hat{n}_{k,\downarrow}\rangle_\sigma(1-\langle\hat{n}_{k,\downarrow}\rangle_\tau),\\
&\langle\hat{n}_{k,+2}\rangle=\langle\hat{n}_{k,\downarrow}\rangle_\sigma)\langle\hat{n}_{k,\downarrow}\rangle_\tau.
\end{aligned}
\end{align}
We can apply the same technique to compute the $k$-point correlation function
$\langle \hat{n}_{j_{1},l_{1}} \dots \hat{n}_{j_{k},l_{k}} \rangle$ by utilizing the existing results for the SSEP \cite{Deriida07,Frassek_2020}.

\section{The asymmetric stochastic process and its exact solutions }\label{sec:asep}\setcounter{equation}{0}

\subsection{Periodic chain}

For the model studied in Sections \ref{sec:periodic} \ref{sec:twist} and \ref{sec:open}, all bulk transition rates are symmetric. In this section, we aim to generalize our model to the asymmetric case.
\paragraph{Integrability}
Let us first introduce the following $R$-matrix
\begin{align}
 \mathsf{R}_{1,2}(u)=
\mathsf{R}^{\sigma}_{\bar 1,\bar 2}(u)\otimes \mathsf{R}^{\tau}_{\tilde 1,\tilde 2}(u) ,\label{RR:xxz} 
\end{align}
where $\mf{R}^\sigma_{\overline{1},\overline{2}}(u)$ and $\mf{R}^\tau_{\widetilde{1},\widetilde{2}}(u)$ are the deformed six-vertex $R$-matrices
\begin{align}
\mf{R}^{\sigma}_{\bar 1,\bar 2}(u)&=\left(
\begin{array}{cccc}
 \sinh (u+\eta_1) & 0 & 0 & 0 \\
 0 & \ee^{-\eta_1 } \sinh u & \ee^{-u} \sinh \eta_1 & 0 \\
 0 & \ee^u \sinh \eta_1 & \ee^{\eta_1 } \sinh u & 0 \\
 0 & 0 & 0 & \sinh (u+\eta_1) \\
\end{array}
\right),\label{Rxxz}\\
\mf{R}^{\tau}_{\tilde1, \tilde 2}(u)&=\mf{R}^{\sigma}_{\bar 1,\bar 2}(u)|_{\eta_1\to \eta_2}.
\end{align}
The $R$-matrix in Eq. (\ref{RR:xxz}) satisfies the YBE
\begin{eqnarray}
\mf{R} _{1,2}(u-v)\mf{R} _{1,3}(u)\mf{R}_{2,3}(v)=\mf{R}_{2,3}(v)\mf{R} _{1,3}(u)\mf{R} _{1,2}(u-v).\label{YBE:xxz}
\end{eqnarray}
\begin{remark}
For $\eta_1 = \eta_2$, the $R$-matrix in Eq. (\ref{RR:xxz}) reduces to that of the $D_2^{(1)}$ algebra \cite{Martins97,Lima-Santos:2003,Nepomechie2017,Li2021} after a simple gauge transformation.
\end{remark} 
Define the transfer matrix 
\begin{align}
\mf{t}(u)={\rm tr}_0\{\mf{R} _{0,N}(u)\mf{R}_{0,N-1}(u)\cdots \mf{R}
_{0,1}(u)\}.\label{tt:xxz}
\end{align}
By using the YBE (\ref{YBE:xxz}) repeatedly, one can prove $[\mf{t}(u),\,\mf{t}(v)]=0$. The logarithm derivative of the transfer matrix $\mf{t}(u)$ yields the transition matrix of a four‑component asymmetric stochastic process under periodic boundary conditions
\begin{align}
\mf{M}&=\left.\frac{\partial\ln\mf{t}(u)}{\partial u}\right|_{u=0}-N(\coth\eta_1+\coth\eta_2)\times \mathbb{I}\no\\
&=\sum_{k=1}^N\mf{M}_{k,k+1},\label{Transition:Matrix:xxz}
\end{align}
where $\mf{M}_{k,k+1}$ describes the particle transitions on sites $k$ and $k+1$ and takes the form
\renewcommand{\arraystretch}{1.5}
\scriptsize
\begin{align}
&\mf{M}_{k,k+1}=\no\\
&\left(
\begin{array}{cccc|cccc|cccc|cccc}
0 & 0 & 0 & 0 & 0 & 0 & 0 & 0 & 0 & 0 & 0 & 0 & 0 & 0 & 0 & 0 \\
0 & -\mathfrak{a}_2 & 0 & 0 & \bar{\mathfrak{a}}_2 & 0 & 0 & 0 & 0 & 0 & 0 & 0 & 0 & 0 & 0 & 0 \\
0 & 0 & -\mathfrak{a}_1 & 0 & 0 & 0 & 0 & 0 & \bar{\mathfrak{a}}_1 & 0 & 0 & 0 & 0 & 0 & 0 & 0 \\
0 & 0 & 0 & -\mathfrak{a}_1\!-\!\mathfrak{a}_2 & 0 & 0 & \bar{\mathfrak{a}}_2 & 0 & 0 & \bar{\mathfrak{a}}_1 & 0 & 0 & 0 & 0 & 0 & 0 \\
\hline 
0 & \mathfrak{a}_2 & 0 & 0 & -\bar{\mathfrak{a}}_2 & 0 & 0 & 0 & 0 & 0 & 0 & 0 & 0 & 0 & 0 & 0 \\
0 & 0 & 0 & 0 & 0 & 0 & 0 & 0 & 0 & 0 & 0 & 0 & 0 & 0 & 0 & 0 \\
0 & 0 & 0 & \mathfrak{a}_2 & 0 & 0 & -\mathfrak{a}_1\!-\!\mathfrak{a}_2 & 0 & 0 & 0 & 0 & 0 & \bar{\mathfrak{a}}_1 & 0 & 0 & 0 \\
0 & 0 & 0 & 0 & 0 & 0 & 0 & -\mathfrak{a}_1 & 0 & 0 & 0 & 0 & 0 & \bar{\mathfrak{a}}_1 & 0 & 0 \\
\hline 
0 & 0 & \mathfrak{a}_1 & 0 & 0 & 0 & 0 & 0 & -\bar{\mathfrak{a}}_1 & 0 & 0 & 0 & 0 & 0 & 0 & 0 \\
0 & 0 & 0 & \mathfrak{a}_1 & 0 & 0 & 0 & 0 & 0 & -\mathfrak{a}_1\!-\!\mathfrak{a}_2 & 0 & 0 & \bar{\mathfrak{a}}_2 & 0 & 0 & 0 \\
0 & 0 & 0 & 0 & 0 & 0 & 0 & 0 & 0 & 0 & 0 & 0 & 0 & 0 & 0 & 0 \\
0 & 0 & 0 & 0 & 0 & 0 & 0 & 0 & 0 & 0 & 0 & -\mathfrak{a}_2 & 0 & 0 & \bar{\mathfrak{a}}_2 & 0 \\
\hline 
0 & 0 & 0 & 0 & 0 & 0 & \mathfrak{a}_1 & 0 & 0 & \mathfrak{a}_2 & 0 & 0 & -\bar{\mathfrak{a}}_1\!-\!\bar{\mathfrak{a}}_2 & 0 & 0 & 0 \\
0 & 0 & 0 & 0 & 0 & 0 & 0 & \mathfrak{a}_1 & 0 & 0 & 0 & 0 & 0 & -\bar{\mathfrak{a}}_1 & 0 & 0 \\
0 & 0 & 0 & 0 & 0 & 0 & 0 & 0 & 0 & 0 & 0 & \mathfrak{a}_2 & 0 & 0 & -\bar{\mathfrak{a}}_2 & 0 \\
0 & 0 & 0 & 0 & 0 & 0 & 0 & 0 & 0 & 0 & 0 & 0 & 0 & 0 & 0 & 0 \\
\end{array}
\right).
\end{align}
\renewcommand{\arraystretch}{1.0}
\normalsize
Here, $\mathfrak{a}_{i}=\coth\eta_i-1$ and $\bar{\mathfrak{a}}_{i}=\coth\eta_i+1$. The transition of the particles follows the same rule as in Section \ref{sec:periodic}, but the transition rates are now \textit{asymmetric}.

Owing to the decomposition property of the $R$-matrix,  the transfer matrix can be written as a product of two sub-transfer matrices
\begin{align}
\mf{t}(u)=\mf{t}_\sigma(u)\otimes \mf{t}_\tau(u),
\end{align} 
where $\mf{t}_\sigma(u)$ and $\mf{t}_\tau(u)$ are defined as 
\begin{align}
\mf{t}_s(u)&={\rm tr}_0\{\mf{T}^s_0(u)\},\quad s=\sigma,\tau,\\
\mf{T}^s_0(u)&=\mf{R} _{0,N}^s(u)\mf{R}_{0,N-1}^s(u)\cdots \mf{R}
_{0,1}^s(u)\no\\
&=\left(
\begin{array}{cc}
\mf{A}_s(u) & \mf{B}_s(u) \\
\mf{C}_s(u) & \mf{D}_s(u) 
\end{array}
\right).
\end{align} 

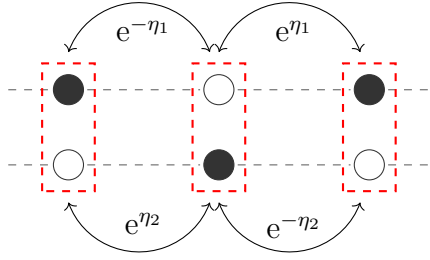
\begin{figure}[htbp]
\centering
\begin{tikzpicture}
\draw[dashed,color=gray,line width=0.5pt] (0.2,0) --(0.75,0);
\draw[dashed,color=gray,line width=0.5pt] (1.25,0) --(2.75,0);
\draw[dashed,color=gray,line width=0.5pt] (3.25,0) --(4.75,0);
\draw[dashed,color=gray,line width=0.5pt] (5.25,0) --(5.8,0);
\draw[color=black!80,fill] (1,0) circle (.2);
\draw[color=black!80] (3,0) circle (.2);
\draw[color=black!80,fill] (5,0) circle (.2);
\draw[<->] (1.0,0.5) arc (160 : 20 : 1.0);
\coordinate[label=above:$\ee^{-\eta_1}$] (1) at (2,0.5);
\draw[<->] (3.0,0.5) arc (160 : 20 : 1.0);
\coordinate[label=above:$\ee^{\eta_1}$] (1) at (4,0.5);
\draw[dashed,color=gray,line width=0.5pt] (0.2,-1) --(0.75,-1);
\draw[dashed,color=gray,line width=0.5pt] (1.25,-1) --(2.75,-1);
\draw[dashed,color=gray,line width=0.5pt] (3.25,-1) --(4.75,-1);
\draw[dashed,color=gray,line width=0.5pt] (5.25,-1) --(5.8,-1);
\draw[color=black!80] (1,-1) circle (.2);
\draw[color=black!80,fill] (3,-1) circle (.2);
\draw[color=black!80] (5,-1) circle (.2);

\draw[<->] (1.0,-1.5) arc (-160 : -20 : 1.0);
\coordinate[label=below:$\ee^{\eta_2}$] (1) at (2,-1.5);
\draw[<->] (3.0,-1.5) arc (-160 : -20 : 1.0);
\coordinate[label=below:$\ee^{-\eta_2}$] (1) at (4,-1.5);
\draw[red,dashed,thick] (0.65,-1.35) rectangle (1.35,0.35);
\draw[red,dashed,thick] (2.65,-1.35) rectangle (3.35,0.35);
\draw[red,dashed,thick] (4.65,-1.35) rectangle (5.35,0.35);
\end{tikzpicture}
\caption{Transitions in the asymmetric stochastic process.}\label{Fig6}
\end{figure}

Taking the first derivative of $\ln\mf{t}_s(u)$ gives the Markov transition matrix of the periodic ASEP
\begin{align}
\mf{M}_\sigma&=\sinh\eta_1\left.\frac{\partial\ln \mf{t}_\sigma(u)}{\partial u}\right|_{u=0}-N\cosh\eta_1\times \mathbb{I}=\sum_{k=1}^N\mf{M}^{(\sigma)}_{k,k+1},\\
\mf{M}_\tau&=\sinh\eta_2\left.\frac{\partial\ln \mf{t}_\tau(u)}{\partial u}\right|_{u=0}-N\cosh\eta_2\times \mathbb{I}=\sum_{k=1}^N\mf{M}^{(\tau)}_{k,k+1},
\end{align} 
where the local transition matrix $\mf{M}^{(s)}_{k,k+1}$ models an ASEP \cite{Essler96,deGier05} and takes the following form in the tensor space  $V_{k}\otimes V_{k+1}$
\begin{align}
\mf{M}^{(\sigma)}_{k,k+1}=
\left(
\begin{array}{cccc}
 0 & 0 & 0 & 0 \\
 0 & -\frac{1}{\mathfrak{q}_1} & \mathfrak{q}_1 & 0 \\
 0 & \frac{1}{\mathfrak{q}_1} & -\mathfrak{q}_1 & 0 \\
 0 & 0 & 0 & 0 \\
\end{array}
\right),\quad \mf{M}^{(\tau)}_{k,k+1}=
\left(
\begin{array}{cccc}
 0 & 0 & 0 & 0 \\
 0 & -\frac{1}{\mathfrak{q}_2} & \mathfrak{q}_2 & 0 \\
 0 & \frac{1}{\mathfrak{q}_2} & -\mathfrak{q}_2 & 0 \\
 0 & 0 & 0 & 0 \\
\end{array}
\right),\quad \mathfrak{q}_i=\ee^{\eta_i}.\label{Local:M:ASEP}
\end{align}
It can be verified that the transition matrix $\mf{M}$ in Eq.  (\ref{Transition:Matrix:xxz}) is a direct sum of $\mf{M}_\sigma$ and $\mf{M}_\tau$
\begin{align}
\mf{M}=\left(\frac{\mf{M}_\sigma}{\sinh\eta_1}\right)\oplus\left(\frac{\mf{M}_\tau}{\sinh\eta_2}\right).
\end{align} 
Let us introduce another matrix 
\begin{align}
\overline{\mf M}=\mf{M}_\sigma\oplus\mf{M}_\tau=\sum_{k=1}^N\overline{\mf{M}}_{k,k+1},
\end{align}
\renewcommand{\arraystretch}{1.5}
where $\overline{\mf{M}}_{k,k+1}$ is given by
\scriptsize
\begin{align}
&\overline{\mf{M}}_{k,k+1}=\no\\
&\left(
\begin{array}{cccc|cccc|cccc|cccc}
 0 & 0 & 0 & 0 & 0 & 0 & 0 & 0 & 0 & 0 & 0 & 0 & 0 & 0 & 0 & 0 \\
 0 & -\frac{1}{\mathfrak{q}_2} & 0 & 0 & \mathfrak{q}_2 & 0 & 0 & 0 & 0 & 0 & 0 & 0 & 0 & 0 & 0 & 0 \\
 0 & 0 & -\frac{1}{\mathfrak{q}_1} & 0 & 0 & 0 & 0 & 0 & \mathfrak{q}_1 & 0 & 0 & 0 & 0 & 0 & 0 & 0 \\
 0 & 0 & 0 & -\frac{1}{\mathfrak{q}_1}\!-\!\frac{1}{\mathfrak{q}_2} & 0 & 0 & \mathfrak{q}_2 & 0 & 0 & \mathfrak{q}_1 & 0 & 0 & 0 & 0 & 0 & 0 \\ [2pt]
\hline
 0 & \frac{1}{\mathfrak{q}_2} & 0 & 0 & -\mathfrak{q}_2 & 0 & 0 & 0 & 0 & 0 & 0 & 0 & 0 & 0 & 0 & 0 \\
 0 & 0 & 0 & 0 & 0 & 0 & 0 & 0 & 0 & 0 & 0 & 0 & 0 & 0 & 0 & 0 \\
 0 & 0 & 0 & \frac{1}{\mathfrak{q}_2} & 0 & 0 & -\frac{1}{\mathfrak{q}_1}\!-\!\mathfrak{q}_2 & 0 & 0 & 0 & 0 & 0 & \mathfrak{q}_1 & 0 & 0 & 0 \\
 0 & 0 & 0 & 0 & 0 & 0 & 0 & -\frac{1}{\mathfrak{q}_1} & 0 & 0 & 0 & 0 & 0 & \mathfrak{q}_1 & 0 & 0 \\ [2pt]
\hline 
 0 & 0 & \frac{1}{\mathfrak{q}_1} & 0 & 0 & 0 & 0 & 0 & -\mathfrak{q}_1 & 0 & 0 & 0 & 0 & 0 & 0 & 0 \\
 0 & 0 & 0 & \frac{1}{\mathfrak{q}_1} & 0 & 0 & 0 & 0 & 0 & -\mathfrak{q}_1\!-\!\frac{1}{\mathfrak{q}_2} & 0 & 0 & \mathfrak{q}_2 & 0 & 0 & 0 \\
 0 & 0 & 0 & 0 & 0 & 0 & 0 & 0 & 0 & 0 & 0 & 0 & 0 & 0 & 0 & 0 \\
 0 & 0 & 0 & 0 & 0 & 0 & 0 & 0 & 0 & 0 & 0 & -\frac{1}{\mathfrak{q}_2} & 0 & 0 & \mathfrak{q}_2 & 0 \\ [2pt]
\hline 
 0 & 0 & 0 & 0 & 0 & 0 & \frac{1}{\mathfrak{q}_1} & 0 & 0 & \frac{1}{\mathfrak{q}_2} & 0 & 0 & -\mathfrak{q}_1\!-\!\mathfrak{q}_2 & 0 & 0 & 0 \\
 0 & 0 & 0 & 0 & 0 & 0 & 0 & \frac{1}{\mathfrak{q}_1} & 0 & 0 & 0 & 0 & 0 & -\mathfrak{q}_1 & 0 & 0 \\
 0 & 0 & 0 & 0 & 0 & 0 & 0 & 0 & 0 & 0 & 0 & \frac{1}{\mathfrak{q}_2} & 0 & 0 & -\mathfrak{q}_2 & 0 \\
 0 & 0 & 0 & 0 & 0 & 0 & 0 & 0 & 0 & 0 & 0 & 0 & 0 & 0 & 0 & 0 \\
\end{array}
\right)
\end{align}
\normalsize
\renewcommand{\arraystretch}{1.0}
One finds that the two matrices $\mf{M}$ and $\overline{\mf{M}}$ are equivalent under a parameter transformation. We will therefore study the exactly solvable model defined by $\overline{\mf{M}}$, rather than the original transition matrix. The model can decompose into two ASEPs with completely independent hopping rates (see Figure \ref{Fig6}) and $\mathcal{Q}_1=n_{+2}+n_{+1}$ and $\mathcal{Q}_2=n_{+2}+n_{-1}$ are still two independent conserved charges.

\paragraph{$T$-$Q$ relation}
The decomposition structure enables direct derivation of the eigenvalues and eigenstates of $\overline{\mf{M}}$ from the known solutions of ${\mf{M}}_s$ and ${\mf{t}}_s(u)$. The eigenvalue of  $\mf{t}_\sigma(u)$, denoted as  $\Lambda_{\sigma}(u)$, can be parameterized by the following $T$-$Q$ relation \cite{Korepin97}
\begin{align} \Lambda_{\sigma}(u)=\ee^{-m\eta_1}\sinh^N(u+\eta_1)\,\frac{Q(u-\eta_1)}{  Q(u)}+\ee^{(N-m)\eta_1}\sinh^N(u)\, \frac{Q(u+\eta_1)}{Q(u)}, \label{TQ:PBC:1}
\end{align}
where $m$ is an integer ranging from 0 to $N$ and the $Q$-function is defined as
\begin{align}
{Q}(u)=\prod_{l=1}^{m}\sinh(u- \mu_l).
\end{align}
The Bethe roots $\{\mu_l\}$ in Eq. \eqref{TQ:PBC:1} should satisfy the BAEs
\bea
\ee^{-N\eta_1}\left[\frac{\sinh(\mu_{k}+\eta_1)}{\sinh(\mu_{k})}\right]^N=
 \prod_{l\ne k}^{m}\frac{\sinh(\mu_k-\mu_l+\eta_1)}
 {\sinh(\mu_k-\mu_l-\eta_1)},\,
 \qquad k=1,\ldots,m.\label{BA:PBC:xxz}
 \eea
Similarly, the eigenvalue of $\mf{t}_{\tau}(u)$, namely $\Lambda_{\tau}(u)$, can be obtained directly from $\Lambda_{\sigma}(u)$ via the substitution
\bea
\{\mu_j,m,\eta_1\}\rightarrow \{ \nu_j,m',\eta_2\}.\eea
The eigenvalue of the
transition matrix $\overline{\mf{M}}$ in terms of the Bethe roots thus reads
\begin{align} E&=\sinh\eta_1\left.\frac{\partial \ln \Lambda_{\sigma}(u)}{\partial u}\right|_{u=0}+\sinh\eta_2\left.\frac{\partial \ln \Lambda_{\tau}(u)}{\partial u}\right|_{u=0}- N\cosh\eta_1-N\cosh\eta_2\no\\
&=\sum_{j=1}^m\frac{\sinh^2\eta_1}{\sinh(\mu_j)\sinh(\mu_j+\eta_1)}+\sum_{j=1}^{m'}\frac{\sinh^2\eta_2}{\sinh(\nu_j)\sinh(\nu_j+\eta_2)}.
\end{align}
\paragraph{Bethe state} The eigenstates of $\mf{t}_\sigma(u)$ and $\mf{t}_\tau(u)$ can be constructed via the algebraic Bethe ansatz approach \cite{Korepin97}
\begin{align}
|\mu_1,\ldots,\mu_m\rangle_\sigma&=\prod_{k=1}^m\mf{B}_\sigma(\mu_k)\binom{1}{0}^{\otimes N}, \\
|\nu_1,\ldots,\nu_{m'}\rangle_\tau&=\prod_{k=1}^{m'}\mf{B}_\tau(\nu_k)\binom{1}{0}^{\otimes N}.
\end{align}
The common eigenstate of $\overline{\mf{M}}$ and $\mf{t}(u)$ thus can be obtained 
\begin{align}
|\mu_1,\ldots,\mu_m\rangle_\sigma\otimes |\nu_1,\ldots,\nu_{m'}\rangle_\tau,
\end{align} 
where $m$ and $m'$ are the eigenvalues of the conserved charges $\mc{Q}_1$ and $\mc{Q}_2$, respectively. 
\subsection{Open chain}
In the asymmetric case, while the system remains integrable under twisted boundary conditions, it is no longer stochastic. In contrast, one can construct a stochastic process with open boundaries. For the $R$-matrix in Eq. (\ref{RR:xxz}), the corresponding reflection matrices are
\begin{align}
\mf{K}^\pm(u)= \mf{K}^{\pm}(\sigma,u) \otimes  \mf{K}^{\pm}(\tau,u),\label{ktk:xxz}
\end{align}
where
\begin{align} 
&\mf{K}^{-}(\sigma,u)=\left(
\begin{array}{cc}
 \ee^{-u} (s_1-s_2) \sinh u-\sinh \eta_1 & - s_2 \sinh (2 u) \\
 - s_1 \sinh (2 u) & \ee^u ( s_2-s_1) \sinh u- \sinh \eta_1 \\
\end{array}
\right), \label{k:xxz:1}\\[4pt]
&\mf{K}^{+}(\sigma,u)= G(\sigma)\mf{K}^{-}(\sigma,-u-\eta_1)|_{\{s_1,s_2\}\rightarrow\{s'_1,s'_2\}}, \quad G(\sigma)={\rm diag}\{\ee^{-\eta_1},\,\ee^{\eta_1}\},\label{k:xxz:2}\\[4pt]
&\mf{K}^{-}(\tau,u)=\left(
\begin{array}{cc}
 \ee^{-u} (t_1-t_2) \sinh u-\sinh \eta_2 & -t_2 \sinh (2 u) \\
 - t_1 \sinh (2 u) & \ee^u ( t_2- t_1) \sinh u-\sinh \eta_2 \\
\end{array}
\right), \label{k:xxz:3}\\[4pt]
&\mf{K}^{+}(\tau,u)= G(\tau)\mf{K}^{-}(\tau,-u-\eta_2)|_{\{t_1,t_2\}\rightarrow\{t'_1,t'_2\}},\quad G(\tau)={\rm diag}\{\ee^{-\eta_2},\,\ee^{\eta_2}\}.\label{k:xxz:4}
\end{align}
The $\mf{K}^{\pm}$-matrices $\mf{K}^{\pm}(\sigma,u)$ and $\mf{K}^{\pm}(\tau,u)$  satisfy the following (dual) reflection equation \cite{Skl88,Cherednik84}
\begin{align}
 &\mf{R}^s _{1,2}(u-v){\mf{K}^{-}_{  1}}(s,u)\mf{R}^s_{2,1}(u+v) {\mf{K}^{-}_{2}}(s,v)\no\\
&={\mf{K}^{-}_{2}}(s,v)\mf{R}^s _{1,2}(u+v){\mf{K}^{-}_{1}}(s,u)\mf{R}^s _{2,1}(u-v),\\
&\mf{R}^s_{1,2}(-u+v){\mf{K}^{ +}_{1}}(s,u)G_1^{-1}(s)\mf{R}^s _{2,1}
 (-u-v-2\eta_s) G_1(s){\mf{K}^{+}_{2}}(s,v)\nonumber\\
&={\mf{K}^{+}_{2}}(s,v)G_1(s)\mf{R}^s _{1,2}(-u-v-2\eta_s) G_1^{-1}(s){\mf{K}^{+}_{1}}(s,u)\mf{R}^s
_{2,1}(-u+v),
 \end{align}
where $\eta_\sigma=\eta_1,\,\eta_\tau=\eta_2$. As in the symmetric case, the $K$-matrices in Eqs. (\ref{k:xxz:1})-(\ref{k:xxz:4}) are not the most general solutions to the reflection equations. Here, we only consider the case where the corresponding transfer matrix can generate a stochastic process.

For the open chain, the transfer matrix is constructed as follows 
\begin{align}
\tilde{\mf{t}}(u)=&\,{\rm tr}_0\{\mf{K}^+_0(u)\mf{R} _{0,N}(u)\mf{R}_{0,N-1}(u)\cdots \mf{R}
_{0,1}(u)\no\\
&\,\times \mf{K}^-_0(u)\mf{R} _{1,0}(u)\mf{R}_{2,0}(u)\cdots \mf{R}
_{N,0}(u)\}.
\end{align}
As in the periodic system, the transfer matrix $\tilde{\mf t}(u)$ is a product of two sub transfer matrices  
\begin{align}
\tilde{\mf{t}}(u)=\tilde{\mf{t}}_\sigma(u)\otimes \tilde{\mf{t}}_\tau(u),
\end{align} 
where $\tilde{\mf{t}}_s(u)$ is defined as 
\begin{align}
\tilde{\mf{t}}_s(u)&={\rm tr}_0\{\mf{K}^+_0(s,u)\mf{R}^s _{0,N}(u)\mf{R}^s_{0,N-1}(u)\cdots \mf{R}^s
_{0,1}(u)\no\\
&\quad \times \mf{K}^-_0(s,u)\mf{R}^s _{N,0}(u)\mf{R}^s_{N-1,0}(u)\cdots \mf{R}^s
_{1,0}(u)\}.
\end{align}
The transfer matrices $\tilde{\mf{t}}_\sigma(u)$ and $\tilde{\mf{t}}_\tau(u)$ each yields the Markov transition matrix of an ASEP with open boundaries
\begin{align}
\widetilde{\mf{M}}_s&=\frac{1}{2}\left.\frac{\partial \ln\bar{\mf{t}}_s(u)}{\partial u}\right|_{u=0}-\mbox{const}\times \mathbb{I}\no\\
&=\sum_{k=1}^{N-1}\mf{M}^{(s)}_{k,k+1}+\mf{M}^{(s)}_{1}+\mf{M}^{(s)}_{N}, \quad s=\sigma,\tau,
\end{align}
where $\mf{M}^{(s)}_{k,k+1}$ is defined in Eq. (\ref{Local:M:ASEP}), and $\mf{M}^{(s)}_{1}$ and $\mf{M}^{(s)}_{N}$ give the transition rules at the left and right boundaries respectively
\begin{align}
&\mf{M}_1^{(\sigma)}=\left(
\begin{array}{cc}
 -s_1 & s_2 \\
 s_1 & -s_2 \\
\end{array}
\right),\quad \mf{M}_N^{(\sigma)}=\left(
\begin{array}{cc}
 s'_1& -s'_2\\
 -s'_1& s'_2 \\
\end{array}
\right),\\
&\mf{M}_1^{(\tau)}=\left(
\begin{array}{cc}
 -t_1 & t_2 \\
 t_1 & -t_2 \\
\end{array}
\right),\quad \mf{M}_N^{(\tau)}=\left(
\begin{array}{cc}
 t'_1& -t'_2\\
 -t'_1& t'_2 \\
\end{array}
\right).
\end{align}
The matrix  $\widetilde{\mf{M}}=\widetilde{\mf{M}}_\sigma+\widetilde{\mf{M}}_\tau$ now serves as the transition matrix describing an asymmetric stochastic process formed by two \textit{independent} open ASEPs. In the following content, we derive the exact solutions of $\widetilde{\mf{M}}$.

\paragraph{$T$-$Q$ relation} 
Define the following functions
\begin{align}
\tilde f(u)=&\,\frac{\sinh(2u+2\eta_1)}
{\sinh(2u+\eta_1)}(-\sinh\eta+ s_1\ee^{-u}\sinh u+s_2\ee^{u}\sinh u)\no\\
&\times (-\sinh\eta_1 - s_2'\ee^{-u}\sinh u-s_1'\ee^{u}\sinh u)\sinh^{2N}(u+\eta_1),\\
\tilde g(u)=&\,\frac{\sinh(2u+2\eta_1)}
{\sinh(2u+\eta_1)}(-\sinh\eta- s_1\ee^{-u}\sinh u-s_2\ee^{u}\sinh u)\no\\
&\times (-\sinh\eta_1 + s_2'\ee^{-u}\sinh u+ s_1'\ee^{u}\sinh u)\sinh^{2N}(u+\eta_1).
\end{align}
The eigenvalue of $\tilde{\mf{t}}_\sigma(u)$, namely  $\tilde\Lambda_{\sigma}(u)$, can be expressed via the following $T$-$Q$ relation 
\begin{align}
&\tilde\Lambda_{\sigma}(u)=
\tilde{f}(u)\frac{\widetilde Q(u-\eta_1)}{\widetilde Q(u)}+\tilde f(-u-\eta_1)\frac{\widetilde Q(u+\eta_1)}{\widetilde Q(u)},
\end{align}
where \bea && \widetilde{Q}(u)=\prod_{l=1}^{N-1}\sinh(u-\mu_l)\sinh(u+\mu_l+\eta_1). \eea
Unlike the periodic chain, the number of Bethe roots should be fixed at $N-1$ \cite{deGier05,Zhang25}.
The Bethe roots $\{\mu_l\}$ should satisfy the Bethe ansatz equations
\begin{align}
\tilde f(\mu_j)\widetilde{Q}(\mu_j-\eta_1)+\tilde f(-\mu_j+\eta_1)\widetilde{Q}(\mu_j+\eta_1)=0,\quad j=1,\ldots,N-1.\label{BAE:xxz:obc}
\end{align}
The eigenvalue of $\widetilde{\mf{M}}_\sigma$ in terms of the Bethe roots is
\begin{align}
E_\sigma=\sum_{j=1}^{N-1}\frac{\sinh^2\eta_1}{\sinh(\mu_j)\sinh(\mu_j+\eta_1)}-(s_1+s_2)+(s'_1+s'_2).\label{en:xxz:obc}
\end{align} 
From the BAEs (\ref{BAE:xxz:obc}), $2^N-1$ eigenvalues of $\widetilde{\mf{M}}_\sigma$ can be obtained, excluding the steady state. Meanwhile, the alternative $T$-$Q$ relation
\begin{align}
\tilde\Lambda_{\sigma}(u)=
 \tilde{g}(u)+\tilde g(-u-\eta_1),
\end{align}
corresponds exclusively to the steady state with $E_\sigma=0$.

The eigenvalue of $\tilde{\mf{t}}_{\tau}(u)$ and $\widetilde{\mf{M}}_\tau$ can be obtained via the following mapping
\begin{align}
&\tilde\Lambda_{\tau}(u)=\tilde\Lambda_{\sigma}(u)|_{\{s_1,s_2,s'_1, s'_2,\mu_j,\eta_1\}\rightarrow \{ t_1,t_2,t'_1, t'_2,\nu_j,\eta_2\}},\\
&E_{\tau}=E_\sigma|_{\{s_1,s_2,s'_1, s'_2,\mu_j,\eta_1\}\rightarrow \{ t_1,t_2,t'_1, t'_2,\nu_j,\eta_2\}}\,\,\mbox{or}\,\, 0.
\end{align}
The eigenvalue of $\widetilde{\mf{M}}$ is a sum of $E_{\tau}$ and $E_\sigma$, and the eigenstate of $\tilde{\mf{t}}(u)$ factorizes into the tensor product of the eigenstates of $\tilde{\mf{t}}_\sigma(u)$ and $\tilde{\mf{t}}_\tau(u)$. The construction of the Bethe-type  eigenstates of $\tilde{\mf{t}}_\sigma(u)$ is complicated, and one has to introduce a set of gauge transformations, see Refs. \cite{cao03,zhang15,ODBA}. Moreover, due to the non-Hermitian nature of $\tilde{\mf{t}}_s(u)$ and $\widetilde{\mf{M}}_s$, the left and right eigenstates are not simply the Hermitian conjugates of each other. We do not demonstrate the Bethe states in this paper.

\section{Conclusion}\label{sec:conclusion}
\setcounter{equation}{0}

In this work, we first construct integrable stochastic processes associated with the $D_2$ algebra, comprising four types of particles distinguished by their charge values and colors. 
We demonstrate that our model can be decomposed into two XXX spin chains (or SSEPs). This decomposition enables us to derive the exact solution of the model directly from the well-established solution of XXX spin chain. Unlike many other high-rank quantum integrable models, we do not need to solve the coupled Bethe ansatz equations for the rapidities \cite{Sutherland75,Essler05,Cao14}. This significantly simplifies the analytical solution of the system. Due to the decomposition property of the spectrum and the eigenstates, the observables can also be calculated analytically. We further examine the asymmetric analogue, which likewise possesses a factorized form and admits decomposition into two ASEPs.
 
The models presented in this paper can be described as two independent lanes where particles are confined to hop within their respective lane, with inter-lane transitions prohibited. This constraint implies that the dynamics of the overall two-lane system decompose into the product of two \textit{independent}, single-lane stochastic processes.

One of our future objectives is to investigate the integrable stochastic process related to the $D_2^{(2)}$ algebra \cite{Reshetikhin:1987,Nepomechie2021,Robertson:2020,Li:2022}. The model corresponding to the $D_2$ algebra can be decomposed into stochastic processes on two independent lanes, the $D_2^{(2)}$ algebra, however, can lead to stochastic processes on two coupled lanes, where particles can hop between two lanes. This suggests the existence of ``two-dimensional" integrable stochastic processes. 
It has been addressed that the transfer matrix of the $D_2^{(2)}$ spin chain can be factorized as the product of
transfer matrices of two staggered XXZ spin chains with fixed spectral difference after a global gauge transformation. In this case, the model more closely resembles a two-lane stochastic processes \cite{popkov2003,mitsudo2005,pronina2006,jiang2008,Schiffmann2010,helms2019}.

This paper presents the first attempt to construct an integrable stochastic process based on a Lie algebra beyond the $A_{n}^{(1)}$ type. We also aim to investigate other high-rank integrable stochastic processes in the future.

\section*{Acknowledgments}

The financial supports from National Key R$\&$D Program of China (Grant No.2021YFA1402104),
the National Natural Science Foundation of China (Grant Nos. 12247103, 12575006, 12305005, 12434006 and 12575007), Major Basic Research Program of Natural Science of Shaanxi Province
(Grant Nos. 2021JCW-19 and 2017ZDJC-32), are gratefully acknowledged.

\appendix 

\end{document}